\documentclass[letterpaper,twocolumn,10pt]{article}
\usepackage{usenix,epsfig,endnotes}
\usepackage{booktabs}
\usepackage{multirow}
\usepackage{enumitem}
\setitemize{topsep=0em,itemsep=-0.5em,leftmargin=1.2em}

\usepackage{siunitx}
\usepackage{graphicx}
\usepackage{url}
\usepackage{balance}
\usepackage{comment}
\usepackage{xspace}
\usepackage{color}
\usepackage[caption=false]{subfig}
\usepackage{lipsum}  

\renewcommand{\paragraph}[1]{\vskip 4pt\noindent\textbf{#1 }}

\newcommand{\ie}{{\it i.e.},~}
\newcommand{\eg}{{\it e.g.},~}
\newcommand{\etal}{{\it et al.}\xspace}
\newcommand{\cf}{{\it c.f.},~}

\newcommand{\lsec}[1]{\label{sec:#1}}
\newcommand{\lfig}[1]{\label{fig:#1}}
\newcommand{\ltab}[1]{\label{tab:#1}}

\newcommand{\rsec}[1]{\S\ref{sec:#1}}
\newcommand{\rfig}[1]{Fig.~\ref{fig:#1}}
\newcommand{\rtab}[1]{Tab.~\ref{tab:#1}}

\newcommand{\tabpreliminary}{
\begin{table}
\small
\centering
\begin{tabular}{lrr}
\toprule
Cause of Loss & Loss Count & Loss Rate\\\midrule
MAC-layer drop & 42 & 4e-3 (0.36\%) \\
Route not found & 32 & 3e-3 (0.27\%) \\
Spurious duplicate & 8 & 7e-4 (0.07\%) \\
Total & 82 & 7e-3 (0.70\%) \\
\bottomrule
\end{tabular}
\caption{Preliminary Study: summary of the losses in a RPL downward routing run.
Total packets sent: 11,730.
}
\ltab{preliminary}
\end{table}
}

\newcommand{\figsprobing}{
\def\figh{3.15cm}
\begin{figure*}[t]
\subfloat[]{
	\includegraphics[height=\figh]{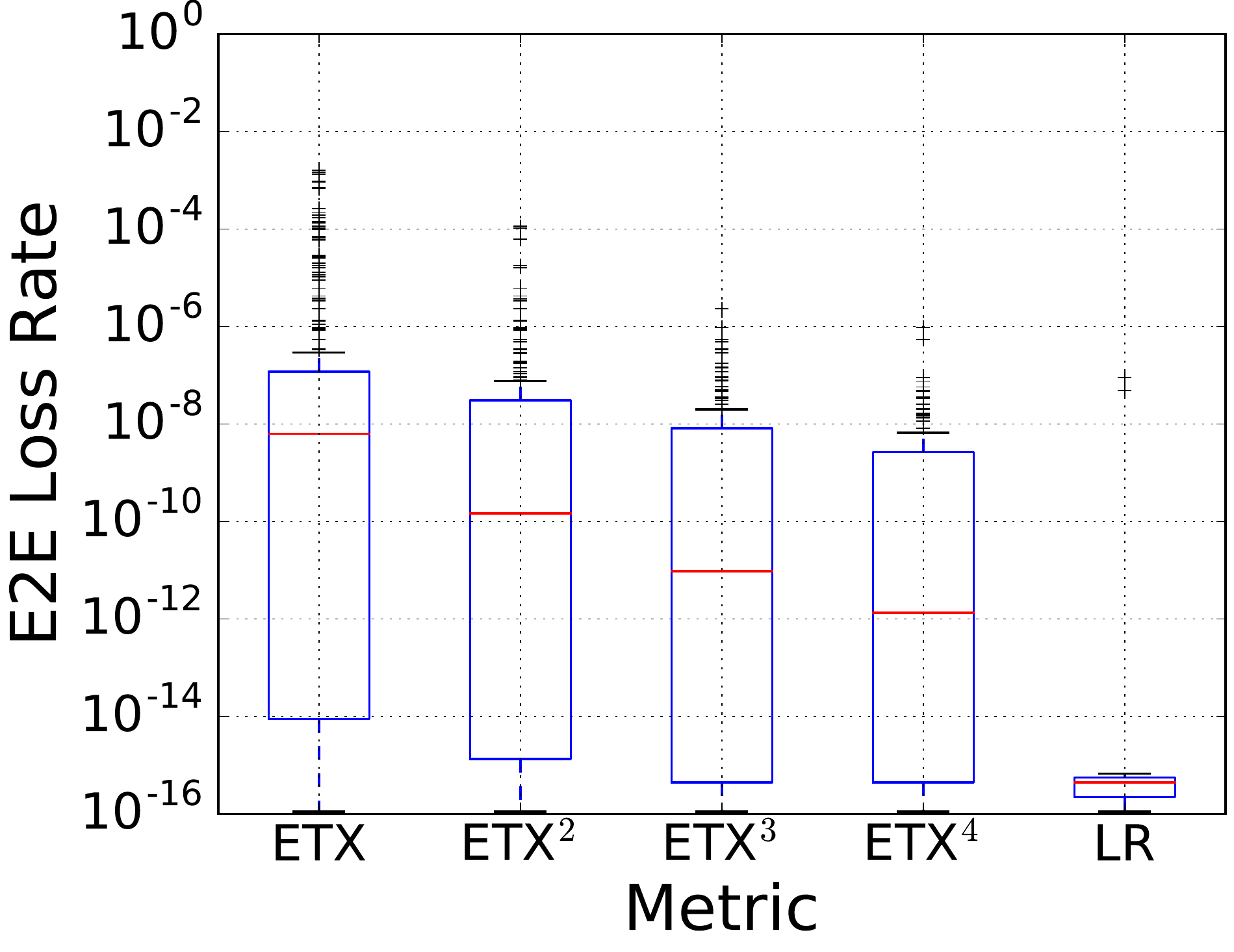}
	\lfig{probing-pdr}
}
\subfloat[]{
	\includegraphics[height=\figh]{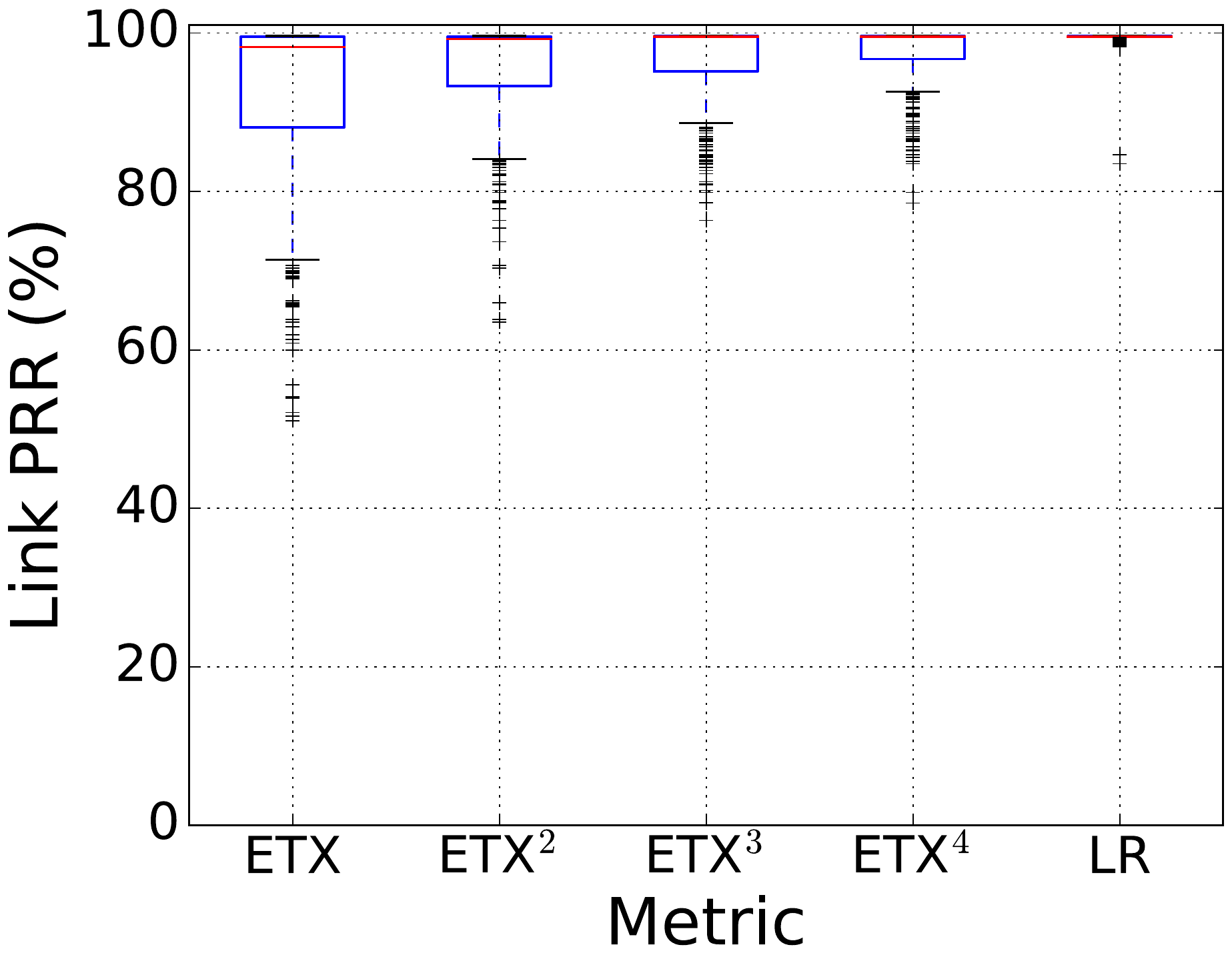}
	\lfig{probing-prr}
}
\subfloat[]{
	\includegraphics[height=\figh]{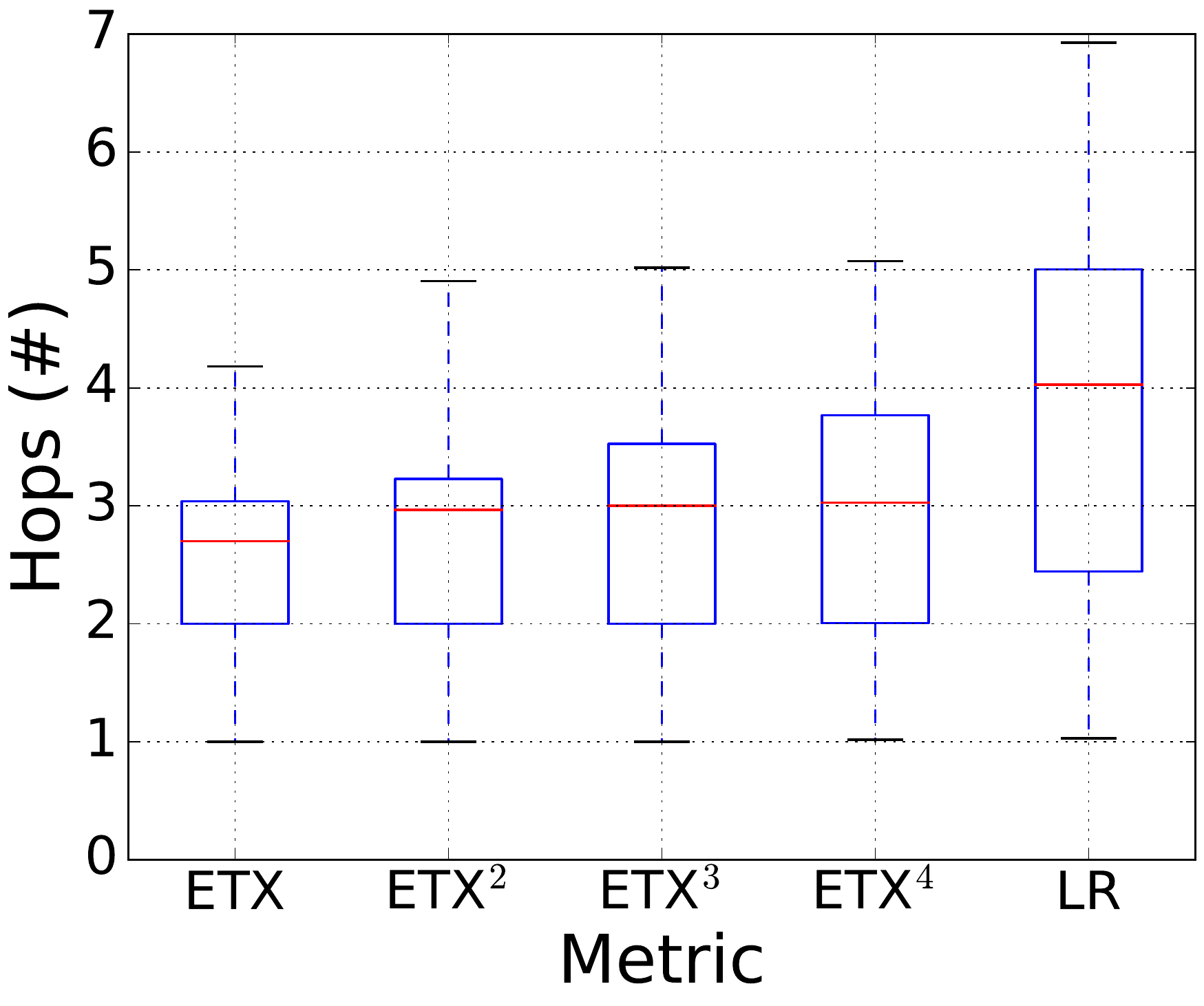}
	\lfig{probing-hops}
}
\subfloat[]{
	\includegraphics[height=\figh]{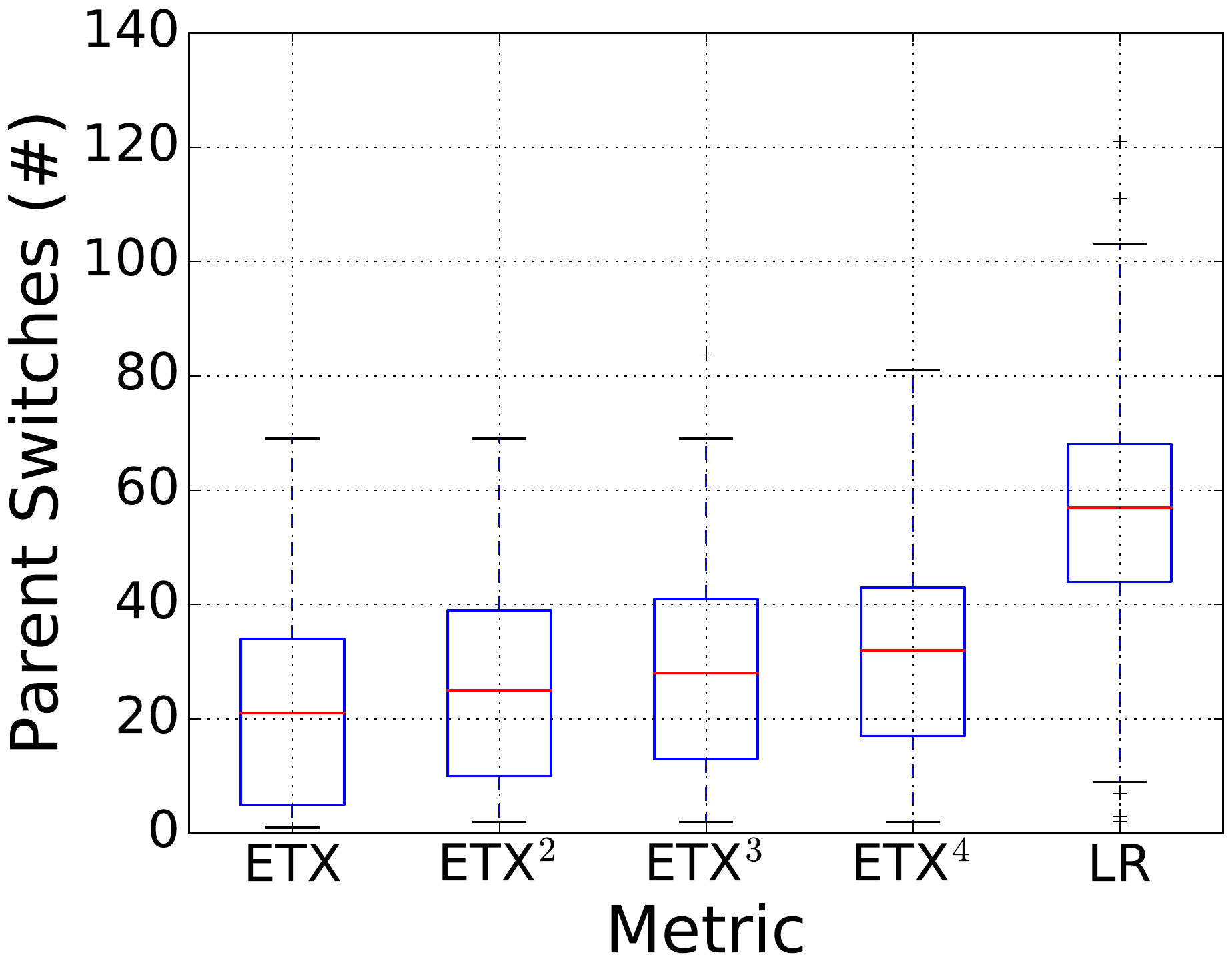}
	\lfig{probing-ps}
}
    \caption{
		Trace-driven simulation: characterization of different link metrics.
		The traditional metric ETX is not best when it comes to end-to-end reliability.
		More reliable metrics, however, result in more hops and parent switches.
    }
    \lfig{probing}
\end{figure*}
}

\newcommand{\figsasym}{
\def\figh{3.44cm}
\begin{figure*}[t]
\centering
\subfloat[]{
	\includegraphics[height=\figh]{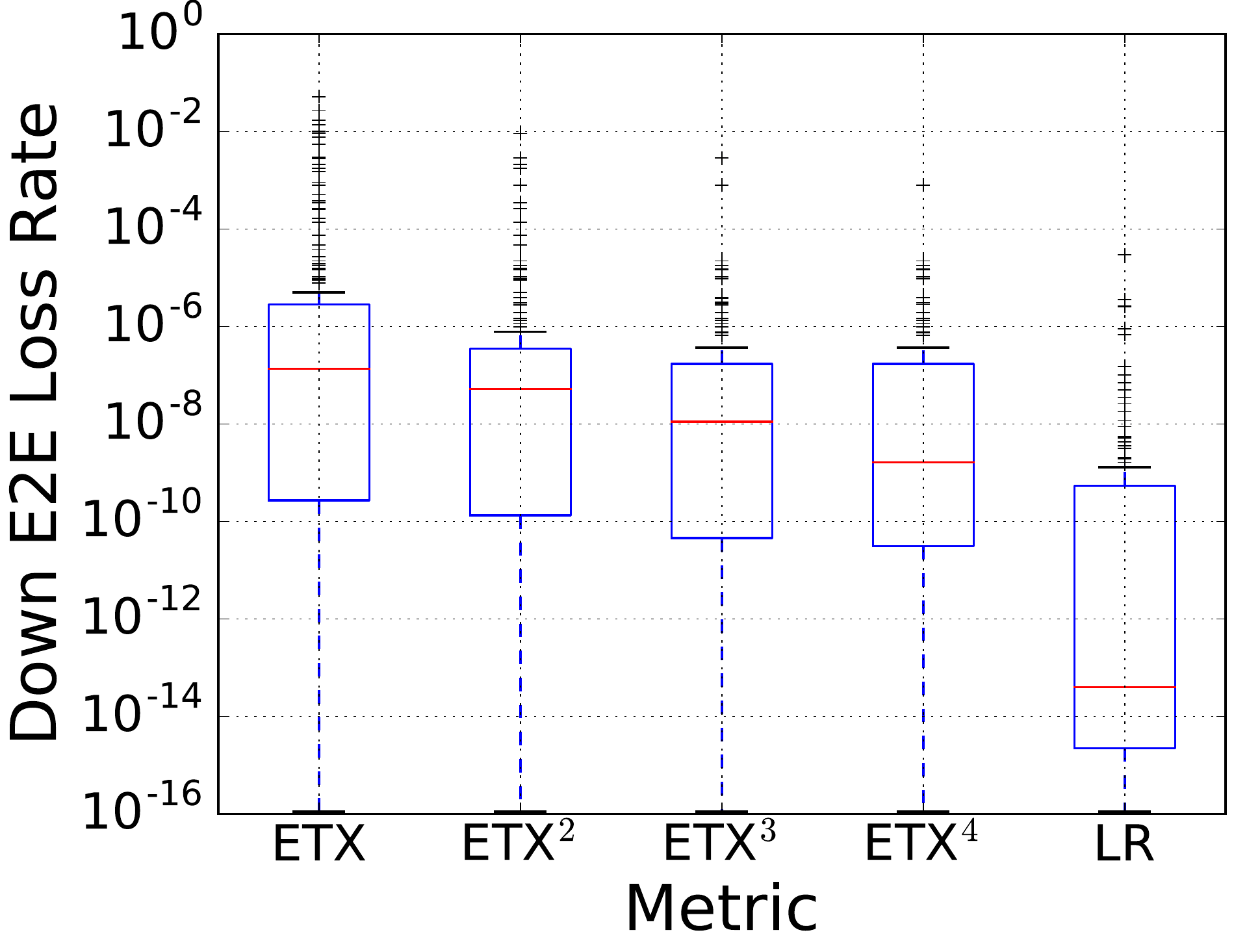}
	\lfig{probing-downpdr}
}
\subfloat[]{
	\includegraphics[height=\figh]{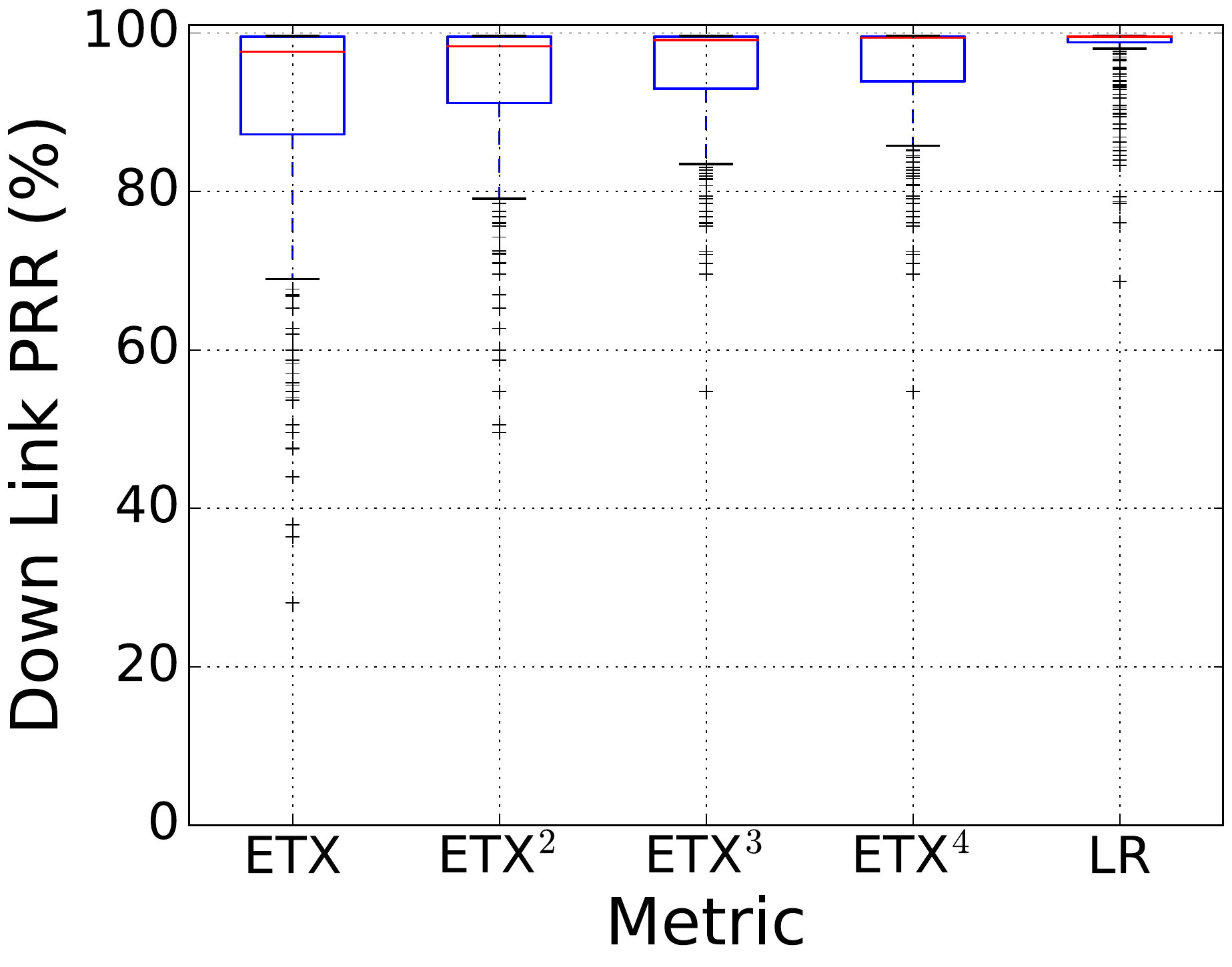}
	\lfig{probing-downprr}
}
\subfloat[]{
	\includegraphics[height=\figh]{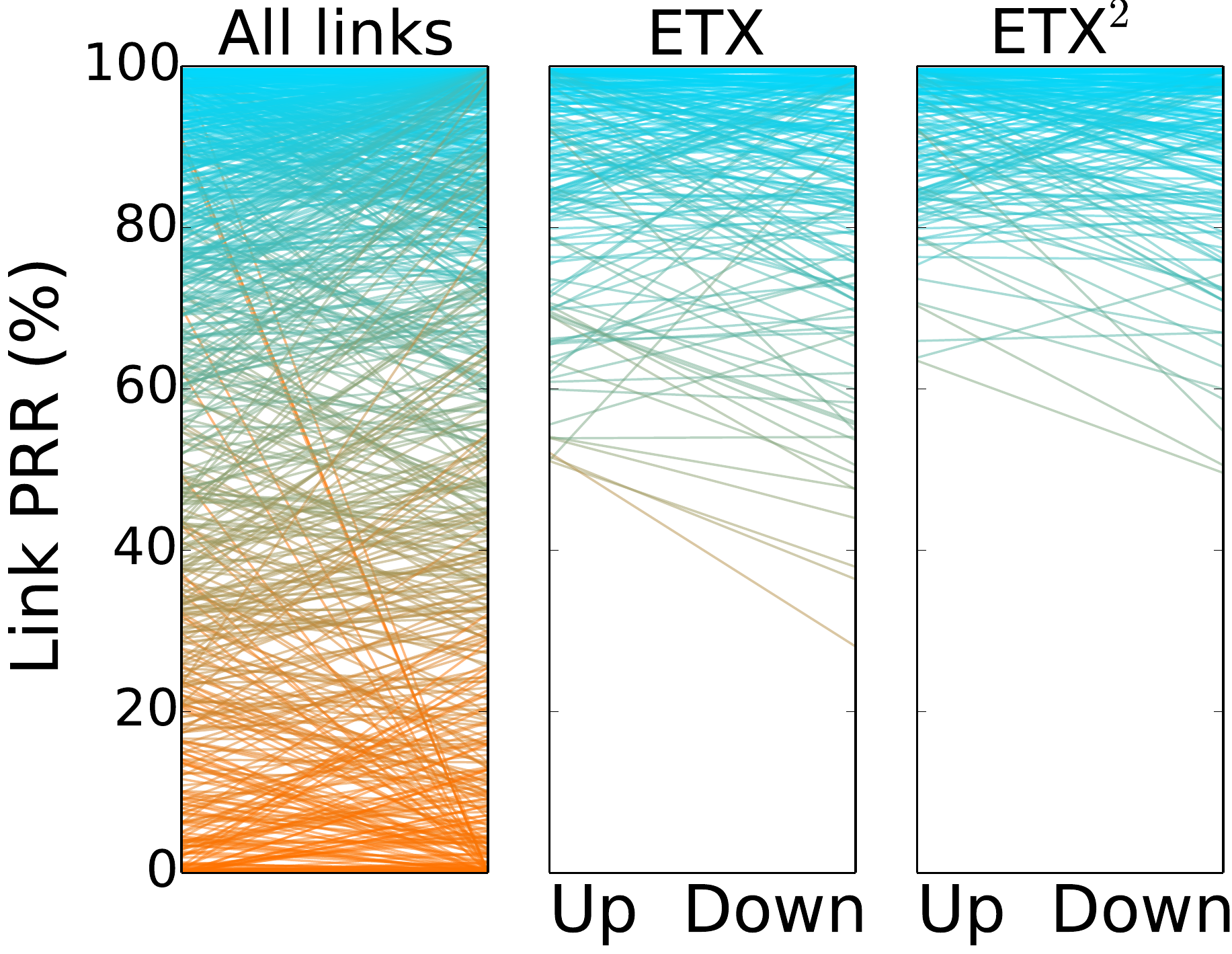}
	\lfig{probing-parallel}
}
    \caption{
		Trace-driven simulation: downward links and asymmetry.
		By simply squaring ETX, higher quality links are selected, improving the worst recorded downward link PRR from 28\% to 50\%.
		Likewise, the worst node's end-to-end loss rate (outlier in \rfig{probing-downpdr}) is brought from 18\% to 3\%.
    }
    \lfig{asym}
\end{figure*}
}

\newcommand{\figstopo}{
\def\figh{2.9cm}
\begin{figure}[t]
\centering
\subfloat[Storing Mode]{
	\includegraphics[height=\figh]{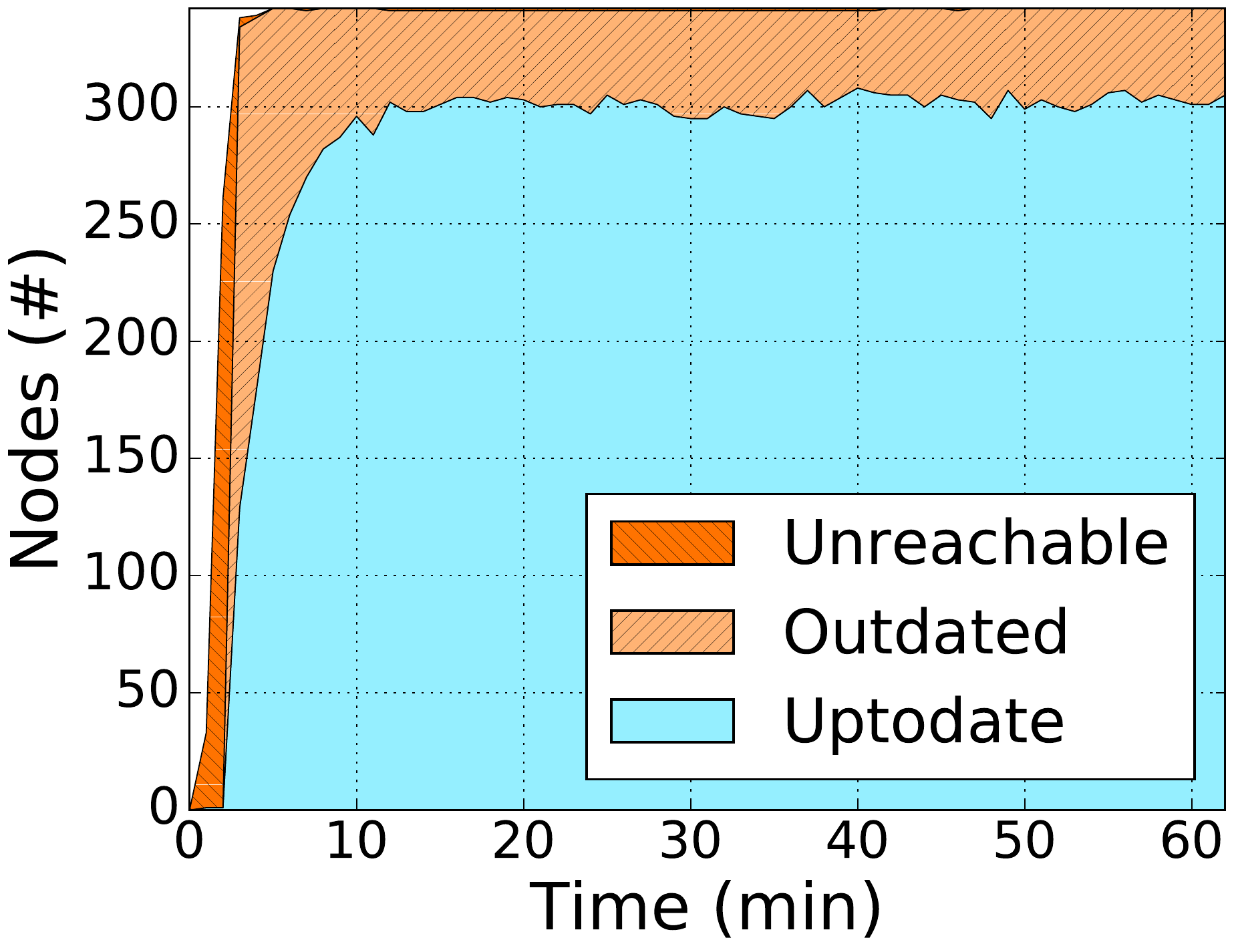}
	\lfig{topology-st}
}
\subfloat[Non-storing Mode]{
	\includegraphics[height=\figh]{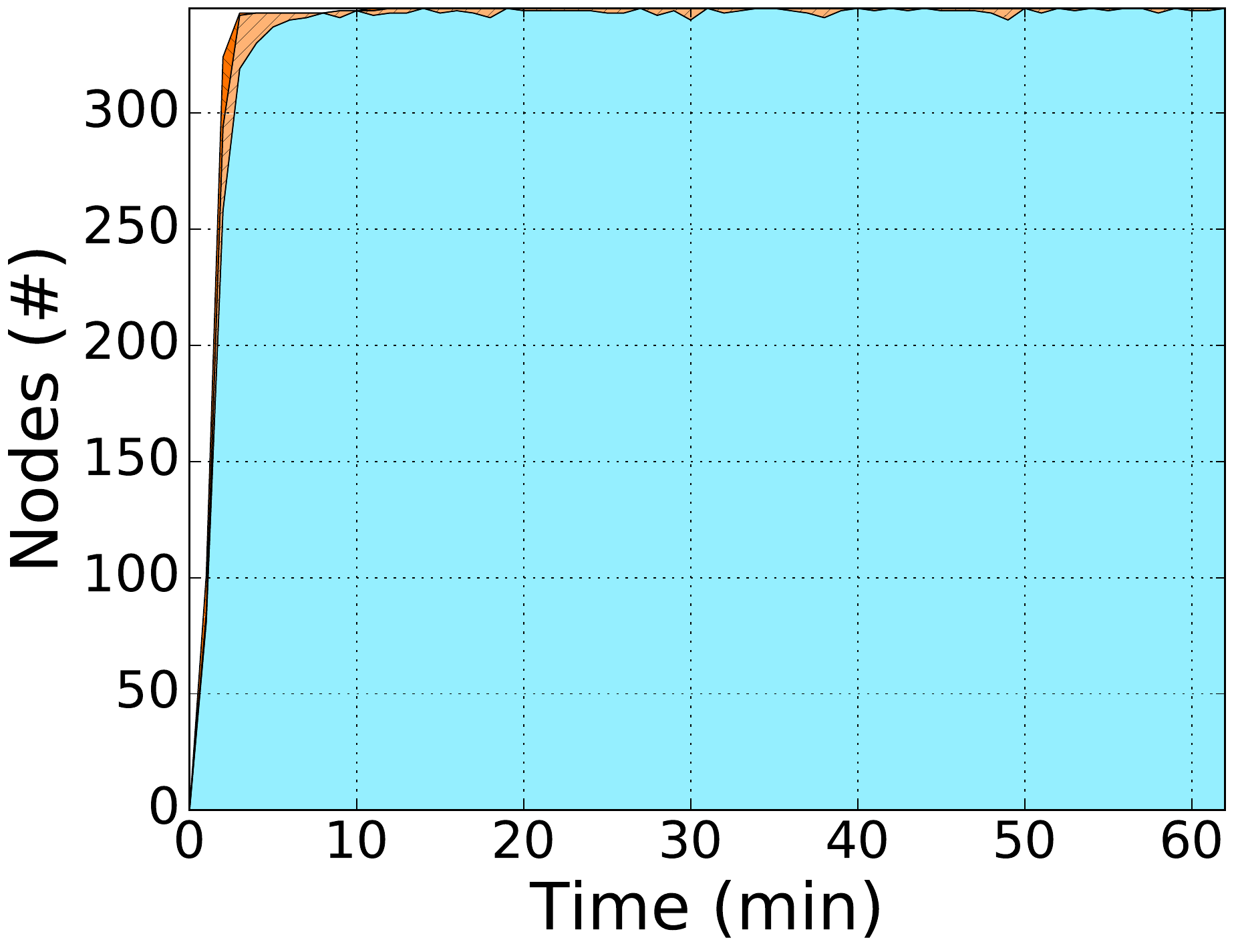}
	\lfig{topology-ns}
}
    \caption{
		Consistency of the routing state.
		Non-storing mode eliminates all inconsistencies after joining, and decreases the portion of outdated routes.
    }
    \lfig{topology}
\end{figure}
}

\newcommand{\fignbrtable}{
\def\figh{3.8cm}
\begin{figure}[t]
\centering
	\includegraphics[height=\figh]{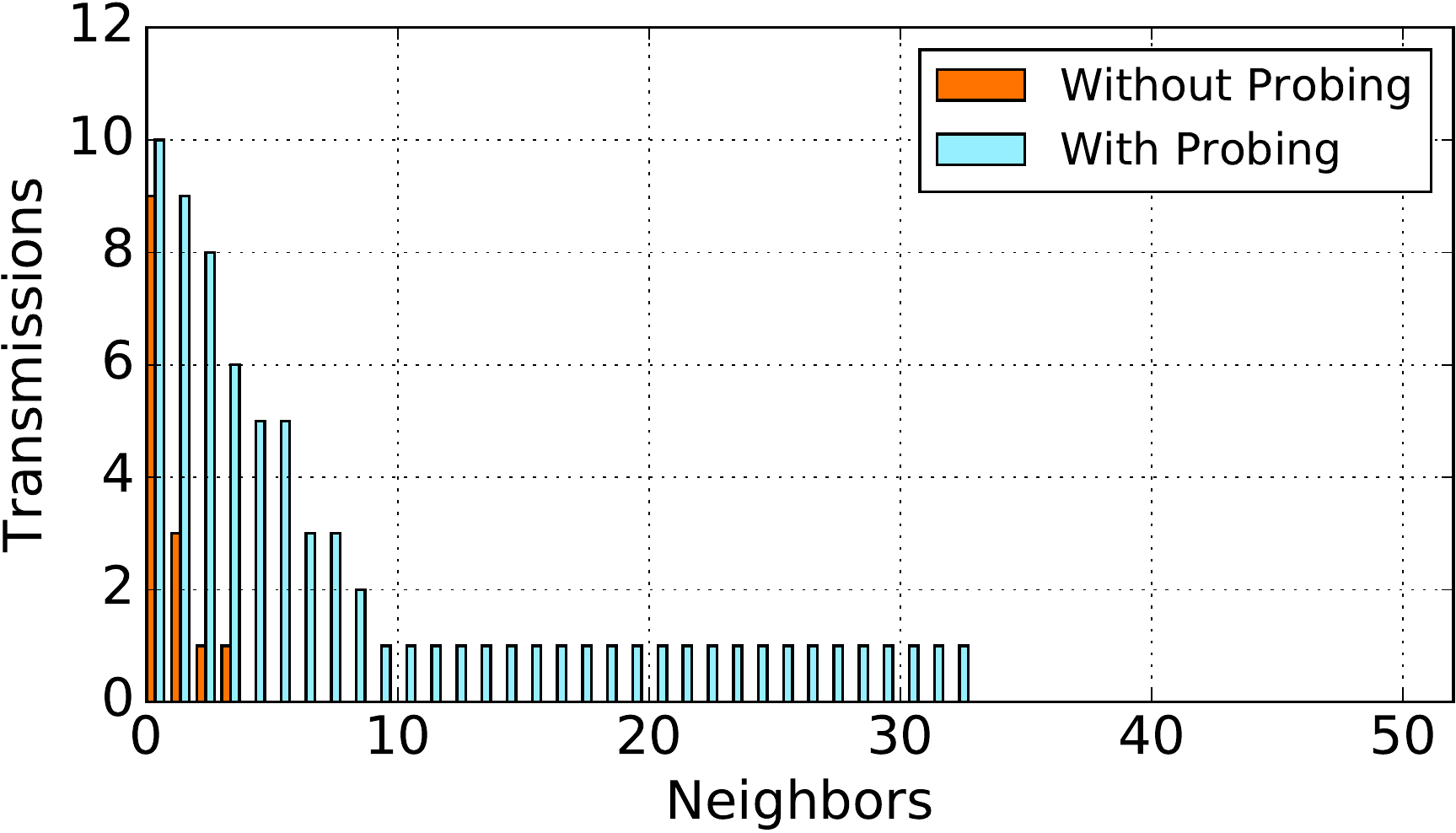}
    \caption{
		Distribution of transmissions over a node's neighbors in a 1h experiment.
		Probing allows to evaluate many more links.
    }
    \lfig{nbrtable}
\end{figure}
}

\newcommand{\figscomparison}{
\def\figh{4cm}
\begin{figure*}[t]
\centering
\subfloat[]{
	\includegraphics[height=\figh]{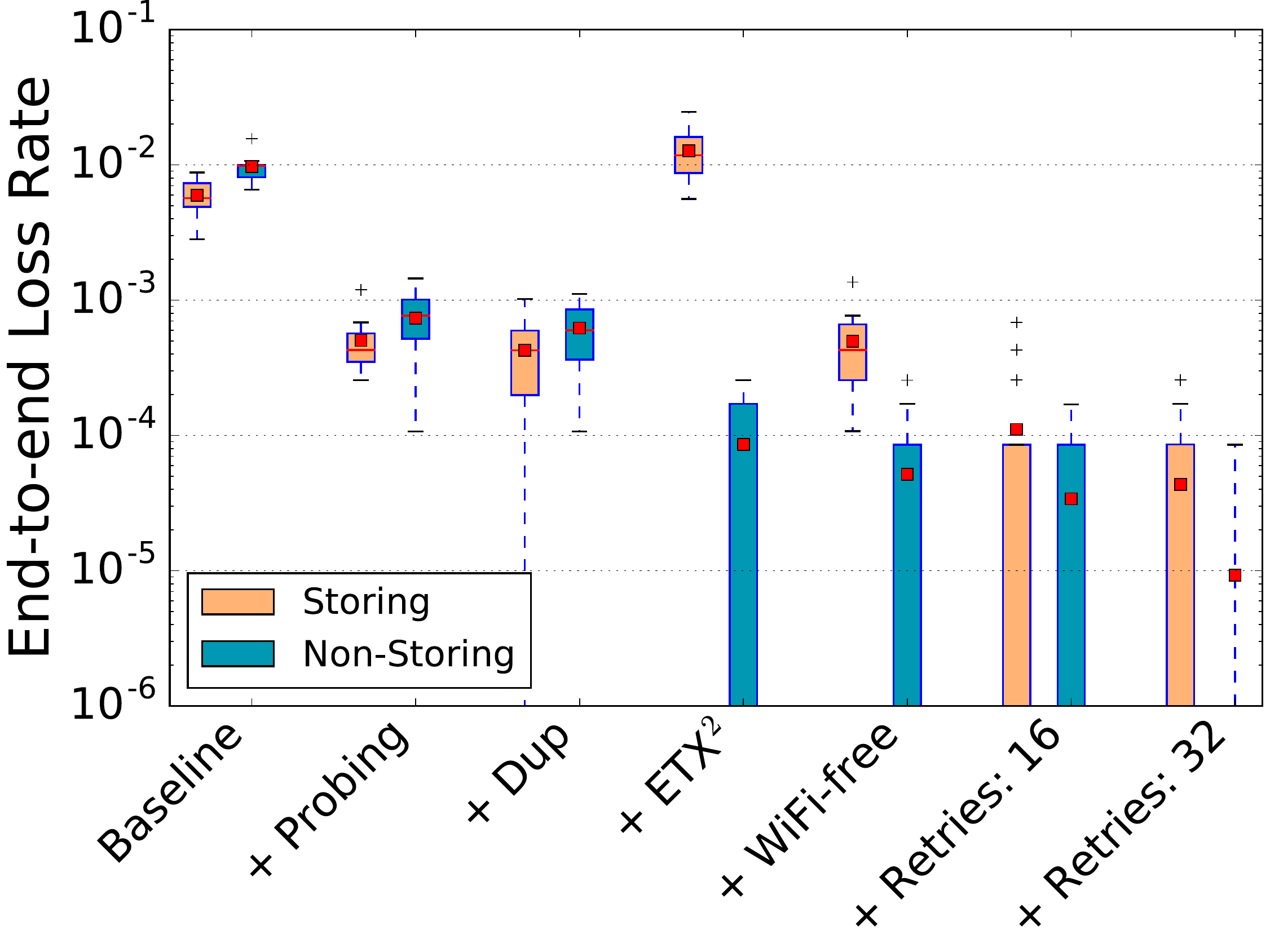}
	\lfig{comparison-pdr}
}
\subfloat[]{
	\includegraphics[height=\figh]{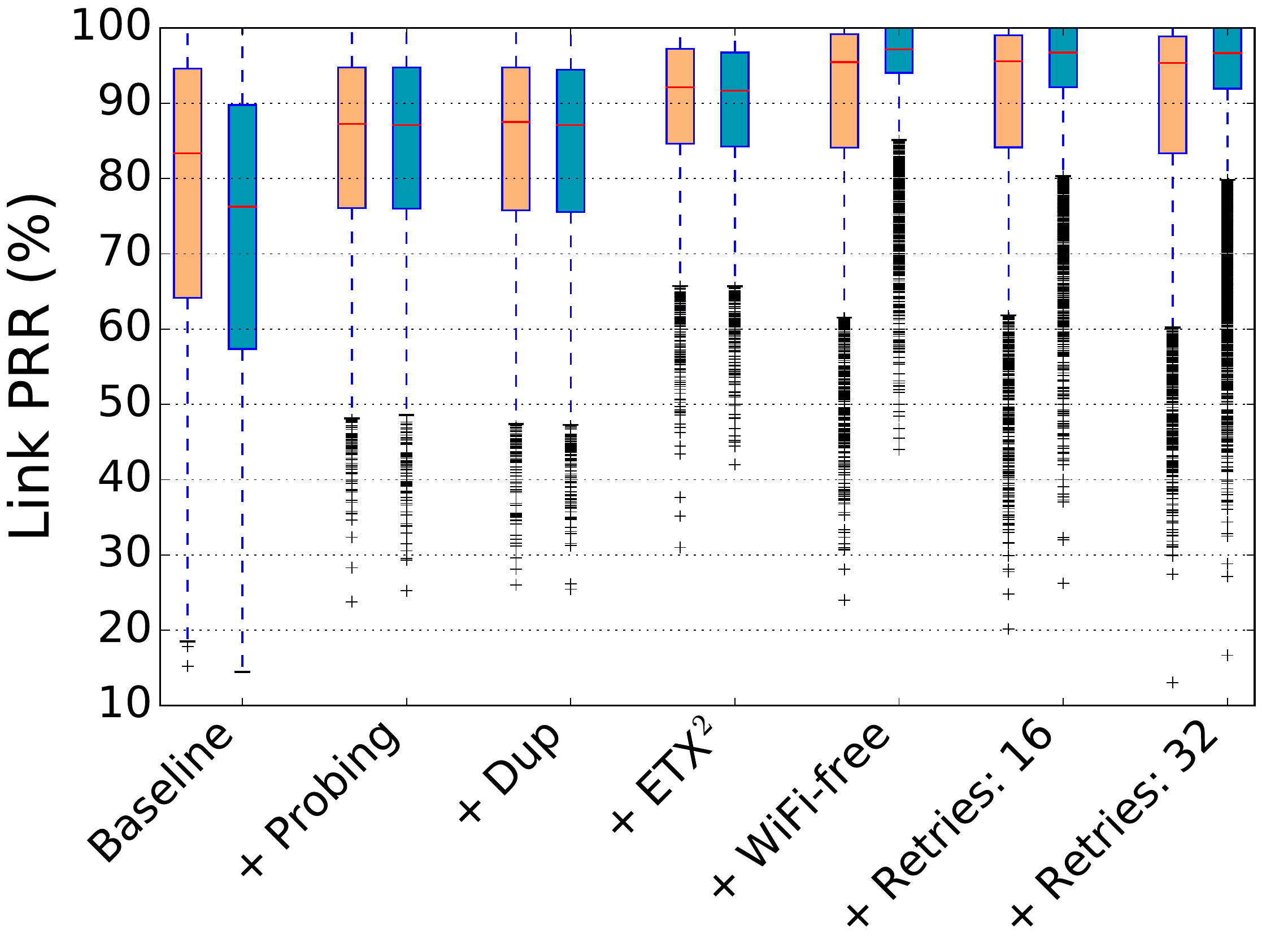}
	\lfig{comparison-prr}
}
\subfloat[]{
	\includegraphics[height=\figh]{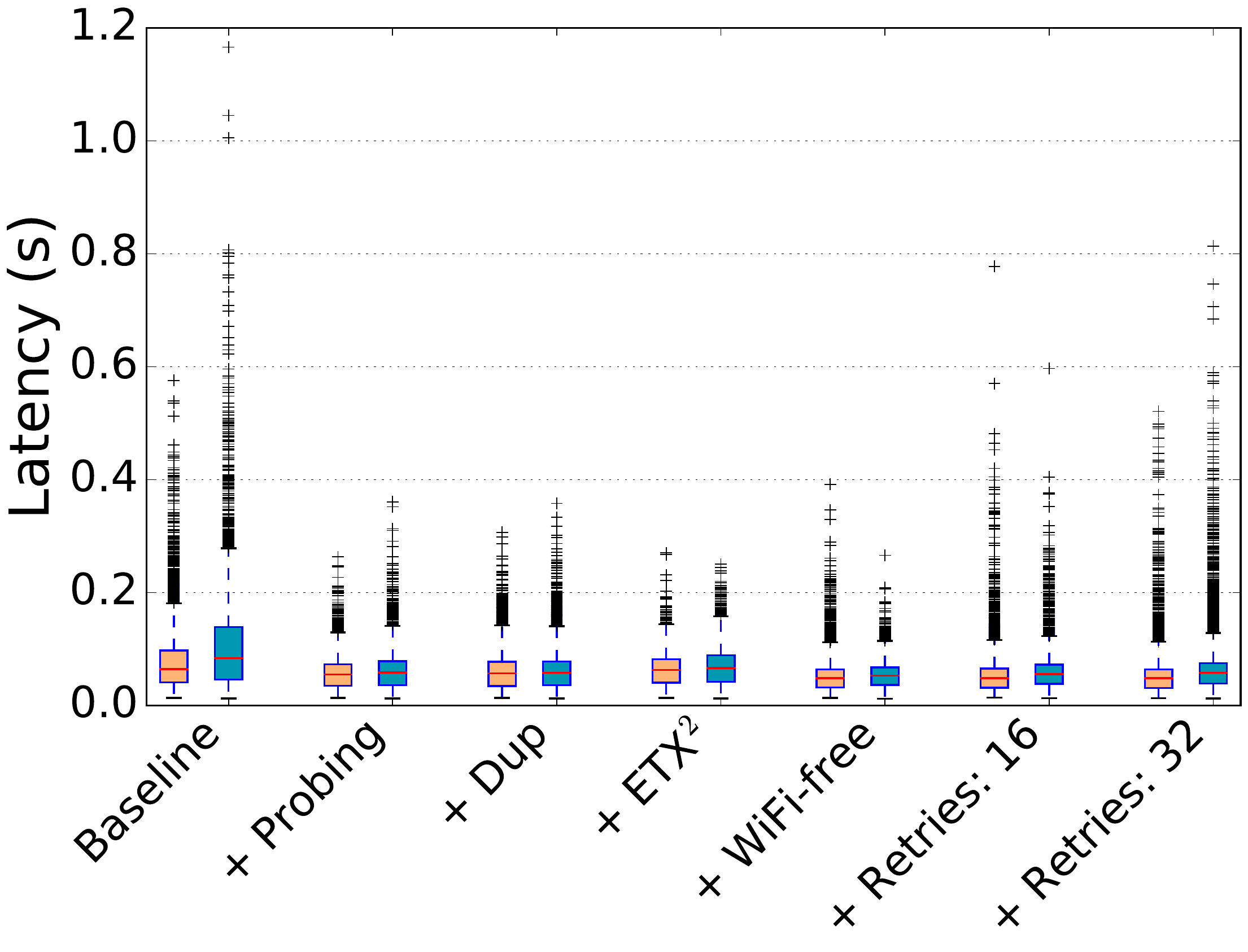}
	\lfig{comparison-latency}
}
    \caption{
		Contribution of the different reliability mechanisms (log y-axis).
		The highest levels of reliability are reached with all mechanisms enabled and in non-storing mode.
		Experiments on WiFi-free channels and with extra retries are done with the best metric, \ie $ETX$ for storing and $ETX^2$ for non-storing.
    }
    \lfig{comparison}
\end{figure*}
}

\newcommand{\figsorchestra}{
\def\figh{3.5cm}
\begin{figure*}[t]
\centering
\subfloat[]{
	\includegraphics[height=\figh]{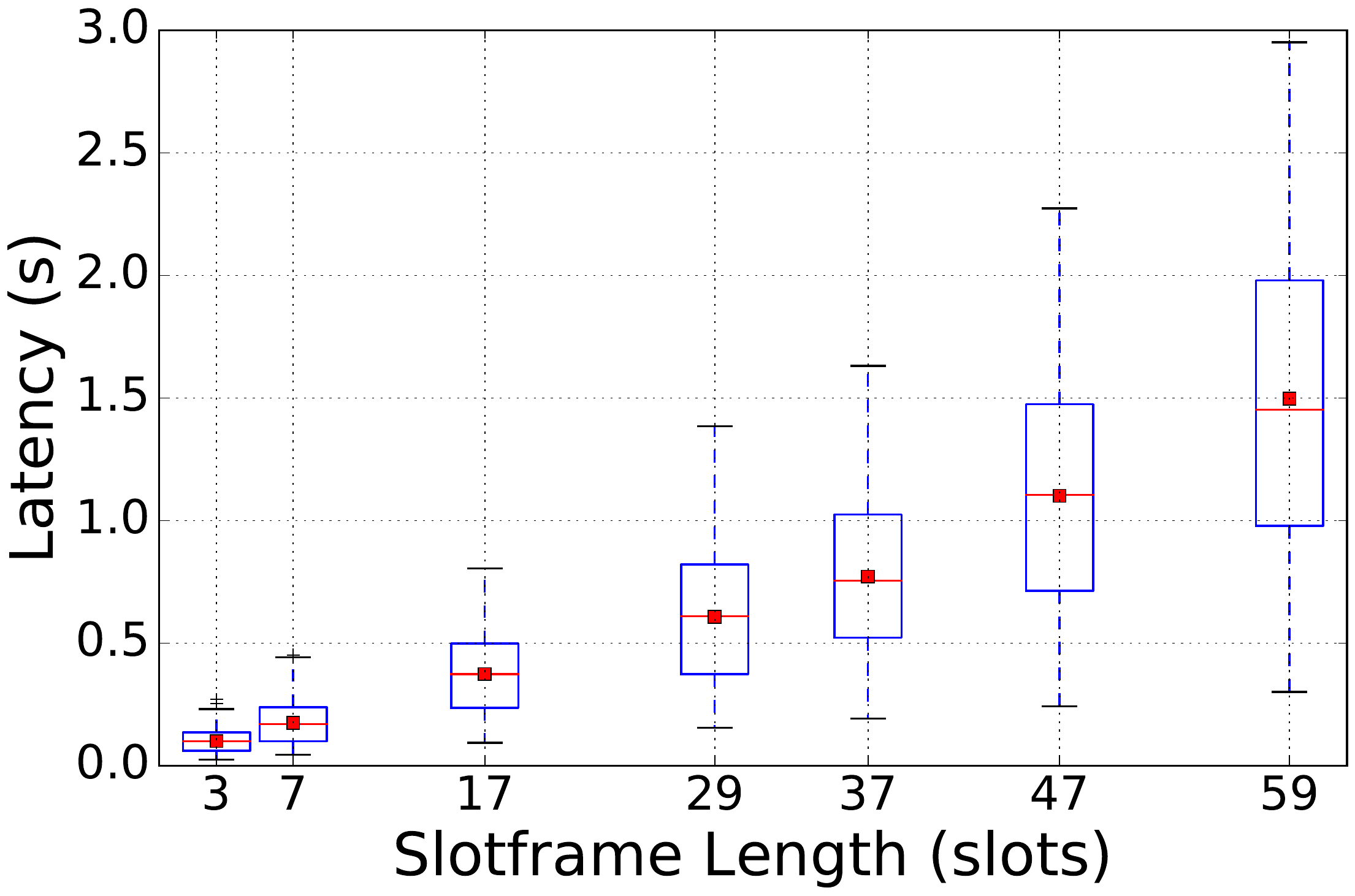}
	\lfig{orchestra-latency}
}
\subfloat[]{
	\includegraphics[height=\figh]{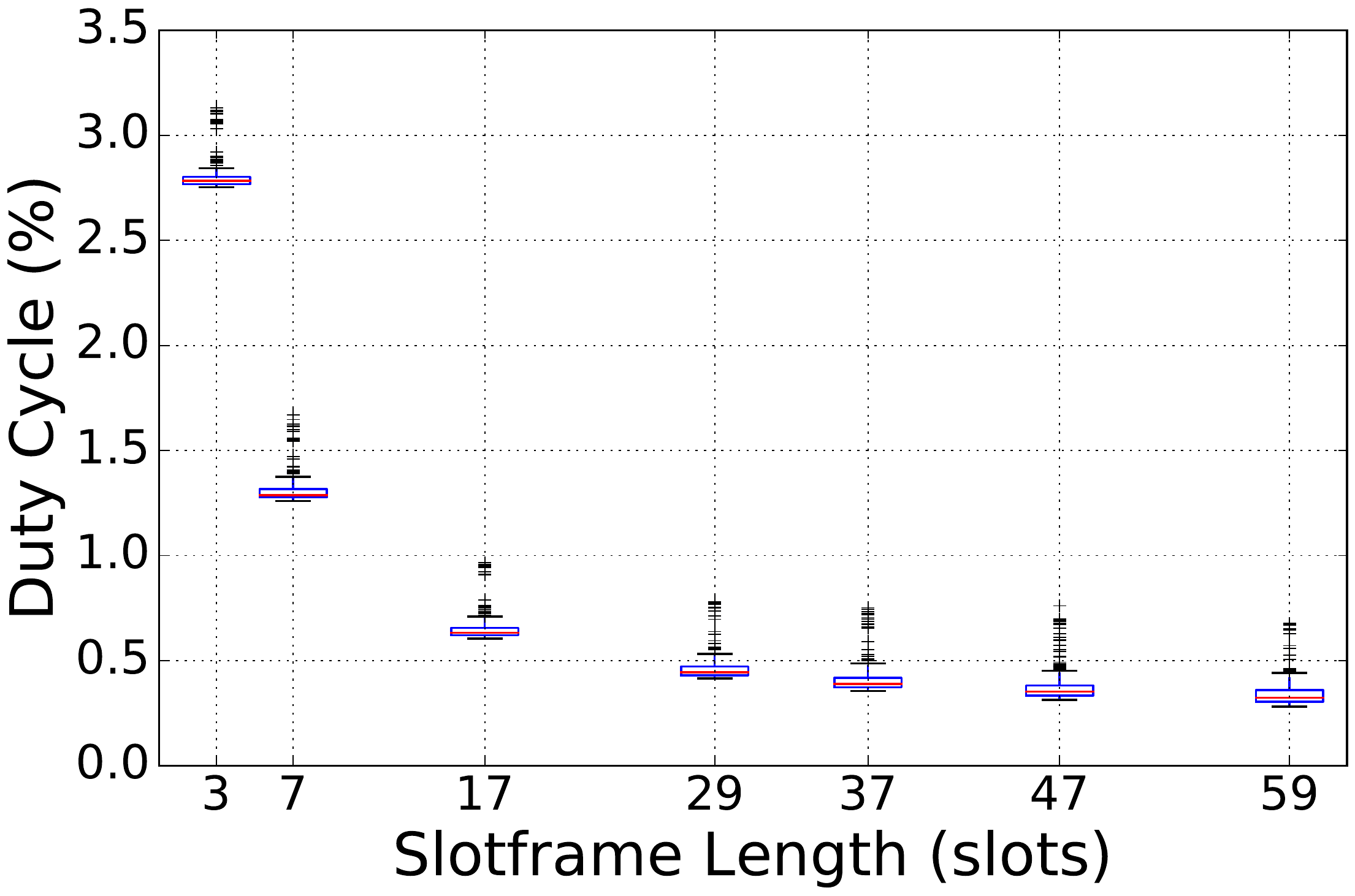}
	\lfig{orchestra-dc}
}
\subfloat[]{
	\includegraphics[height=\figh]{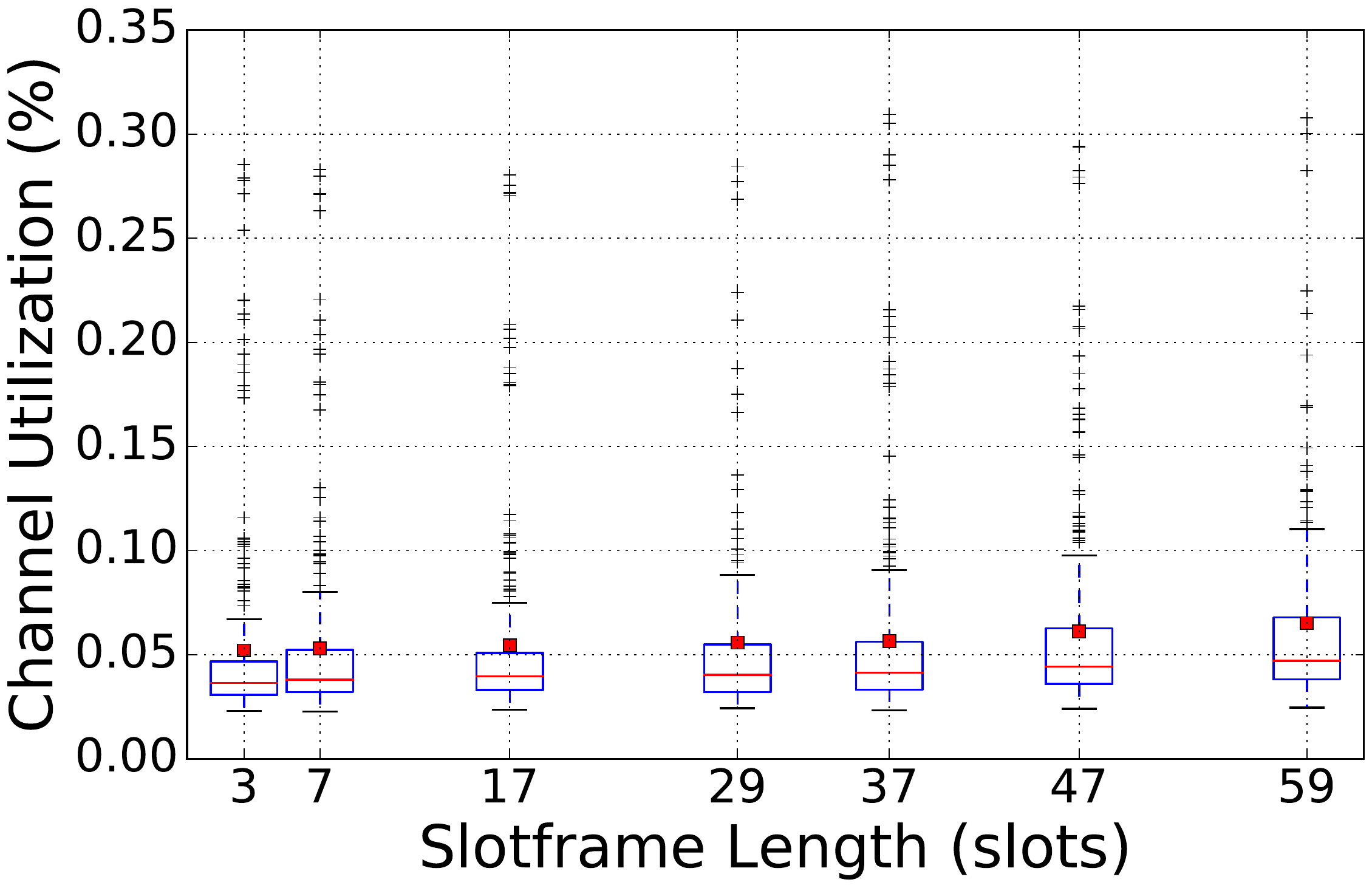}
	\lfig{orchestra-dctx}
}
    \caption{
		Low-power settings.
		Results with Orchestra in the 52-node Grenoble testbed.
		There is a basic trade-off between latency and duty cycle, but most settings achieve sub-second latency and sub-percent duty cycle.
		The channel utilization at every node remains extremely low, below 0.07\% (shared over 16~channels).
    }
    \lfig{orchestra}
\end{figure*}
}

\newcommand{\figtraffic}{
\begin{figure}[t]
\centering
	\includegraphics[width=\columnwidth]{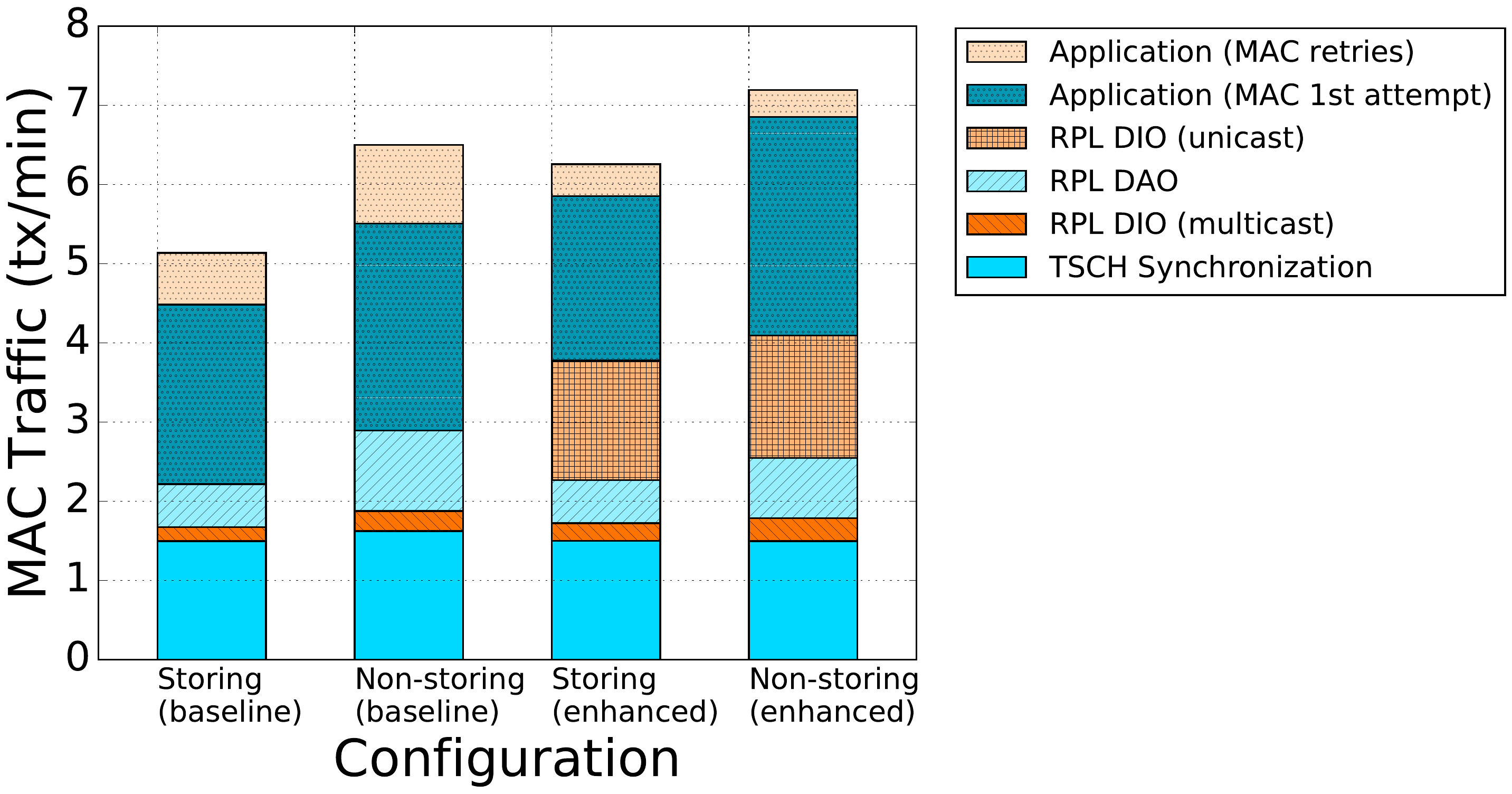}
    \caption{
		Traffic Overhead.
		Our reliability mechanisms increase traffic by 13--23\%.
    }
    \lfig{traffic}
\end{figure}
}

\newcommand{\tabsummary}{
\begin{table*}
\centering
\footnotesize
\begin{tabular}{l@{\hskip 5pt}l@{\hskip 5pt}r@{\hskip 5pt}r@{\hskip 5pt}rlrrrrr}
\toprule
\multicolumn{7}{c}{Setup} & \multicolumn{4}{c}{Loss rate} \\ \cmidrule(r){1-7}\cmidrule(l){8-11}
Testbed & Node & Size & Density\mbox{*} & Radius\mbox{**} & Configuration & \#packets & MAC & Route & Dup & {\bf Total}
 \\\midrule
IoT-LAB Gre. & M3 & 352 & 72 & 6.7 & Storing {\em(baseline)}                               & 117K & 3e-3 & 3e-3 & 4e-4 & {\bf 6e-3}\\
 & & & & & Non-storing {\em(baseline)}                                                     & 117K & 9e-3 & 0 & 9e-4 & {\bf 1e-2}\\
 & & & & & Storing                                                                         & 151K & 4e-4 & 5e-5 & 0 & {\bf 4e-4}\\
 & & & & & Non-storing                                                                     & 157K & 9e-5 & 0 & 0 & {\bf 9e-5}\\
 & & & & & Storing {\em(Wifi-free, 32 rtx)}                                                & 227K & 9e-6 & 3e-5 & 0 & {\bf 4e-5}\\
 & & & & & Non-storing {\em(Wifi-free, 32 rtx)}                                            & 585K & 8e-6 & 0 & 0 & {\bf 8e-6}\\\midrule
IoT-LAB Gre.-52 & M3 & 52 & 8.4 & 5.9 & Non-storing {\em(baseline)}                        & 131K & 8e-2 & 0 & 0 & {\bf 8e-2}\\
 & & & & & Non-storing                                                                     & 606K & 2e-5 & 0 & 0 & {\bf 2e-5}\\
 & & & & & Non-storing {\em(Orchestra)}                                                         & 608K & 3e-5 & 0 & 0 & {\bf 3e-5}\\
 & & & & & Non-storing {\em(Wifi-free, 32 rtx)}                                            & 762K & 0 & 0 & 0 & {\bf 0}\\\midrule
IoT-LAB Lille & M3  & 240 & 237 & 2.4 & Non-storing {\em(baseline)}                         &  35K & 7e-4 & 0 & 0 & {\bf 7e-4}\\
 & & & & & Non-storing                                                                      & 103K & 8e-5 & 0 & 0 & {\bf 8e-5}\\
 & & & & & Non-storing {\em(Wifi-free, 32 rtx)}                                             & 522K & 0 & 0 & 0 & {\bf 0}\\\midrule
Flocklab & OpenMote & 9 & 5.4 & 3.9 & Non-storing {\em(baseline)}                           &  82K & 8e-1 & 0 & 0 & {\bf 8e-1}\\
 & & & & & Non-storing                                                                      & 179K & 5e-5 & 0 & 0 & {\bf 5e-5}\\
 & & & & & Non-storing {\em(Wifi-free, 32 rtx)}                                             & 584K & 2e-5 & 0 & 0 & {\bf 2e-5}\\\midrule
JN-IoT & JN5168 & 24 & 16 & 3.8 & Non-storing {\em(baseline)}                               & 128K & 1e-2 & 0 & 0 & {\bf 1e-2}\\
 & & & & & Non-storing                                                                      & 166K & 4e-4 & 0 & 0 & {\bf 4e-4}\\
 & & & & & Non-storing {\em(Wifi-free, 32 rtx)}                                             & 371K & 2e-5 & 0 & 0 & {\bf 2e-5}\\
\bottomrule
\end{tabular}
\\\hfill\mbox{*} Density is computed as the mean number of items in the nodes' neighbor table.
\\\hfill\mbox{**} Radius is computed as the mean hop count of the farthest away node from the root.
\caption{Downward-traffic experiments summary.
With our reliability mechanisms enabled, we reach loss rates between 2e-5 and 0 (PDR between 99.998\% and no loss observed) across four different testbeds.
}
\ltab{summary}
\end{table*}
}

\newcommand{\tabatoa}{
\begin{table*}
\centering
\footnotesize
\begin{tabular}{lrrrrrrr}
\toprule
\multirow{2}{*}{Configuration}  & \multicolumn{3}{c}{Path} & \multicolumn{4}{c}{Loss rates} \\ \cmidrule(lr){2-4} \cmidrule(l){5-8}
 & Hops & Latency (s) & Root-free paths (\%) & MAC & Route & Dup & {\bf Total} \\\midrule
Storing {\em(ETX)}                              & 5.5 & 0.14 & 13.7 & 2e-3 & 2e-2 & 0 & {\bf 2e-2}\\
Non-storing {\em(ETX)}                          & 5.8 & 0.16 & 0.8 & 1e-3 & 0 & 0 & {\bf 1e-3}\\
Storing {\em(ETX$^2$)}                          & 6.7 & 0.16 & 22.8 & 1e-3 & 5e-2 & 0 & {\bf 5e-2}\\
Non-storing {\em(ETX$^2$)}                      & 7.9 & 0.20 & 1.2 & 6e-4 & 0 & 0 & {\bf 6e-4}\\\midrule
Storing {\em(ETX, Wifi-free, 32 rtx)}           & 5.1 & 0.11 & 12.9 & 4e-5 & 5e-3 & 0 & {\bf 5e-3}\\
Non-storing {\em(ETX$^2$, Wifi-free, 32 rtx)}   & 7.3 & 0.15 & 1.1 & 1e-5 & 0 & 0 & {\bf 1e-5}\\
\bottomrule
\end{tabular}
\caption{Point-to-point routing experiment.
Storing mode suffers from more routing inconsistency than in downward-only experiments, while non-storing remains immune to the problem.
Non-storing mode, however, results in longer paths, as nearly all packets are routed via the root.
ETX$^2$ further increases path length.
}
\ltab{atoa}
\end{table*}
}

\newcommand{\figlongtimeline}{
\begin{figure*}[t]
\centering
	\includegraphics[width=\textwidth]{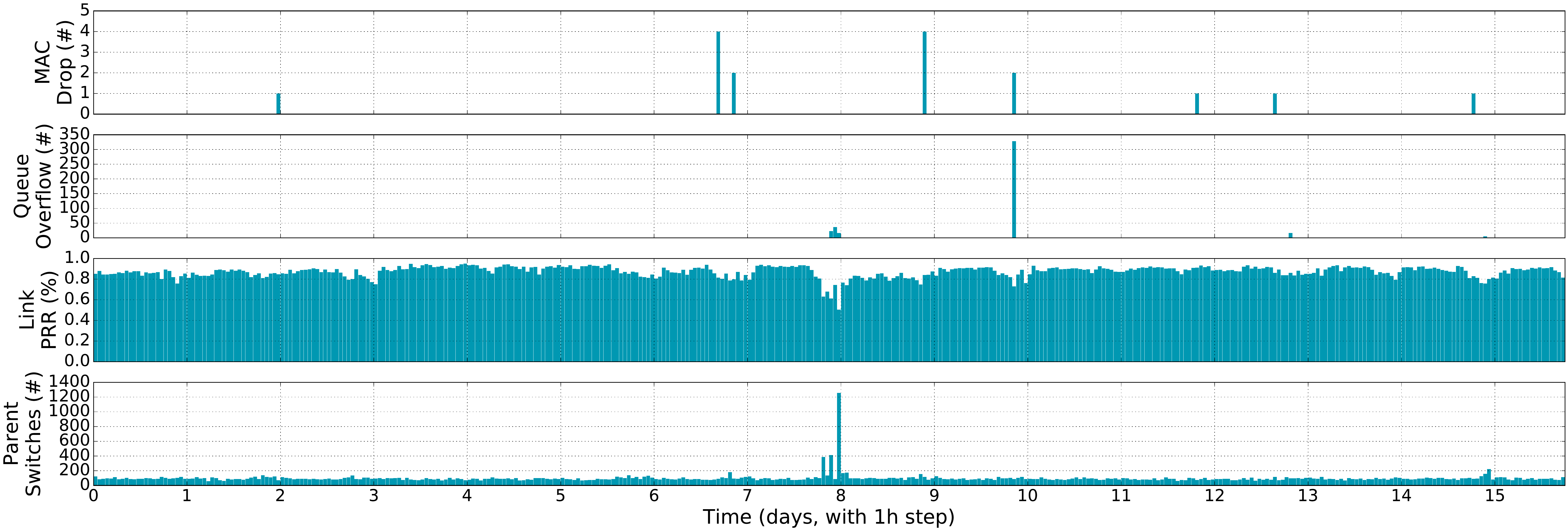}
    \caption{
		Overview of the 2-week experiment in the 24-node JN-IoT testbed.
		Our system operates reliably overall, hiding the fluctuations in link condition.
		Queue overflow rate: 8e-5, MAC drop rate: 3e-6.
    }
    \lfig{longtimeline}
\end{figure*}
}

\begin{document}

\date{}

\title{\Large \bf Five-Nines Reliable Downward Routing in RPL}

\author{
{\rm Simon Duquennoy}\\
simon.duquennoy@ri.se\\
RISE SICS, Sweden and\\Inria Lille -- Nord Europe, France
\and
{\rm Joakim Eriksson}\\
joakim.eriksson@ri.se\\
RISE SICS, Sweden
 \and
 {\rm Thiemo Voigt}\\
thiemo.voigt@ri.se\\
RISE SICS and\\Uppsala University, Sweden
} 

\maketitle

\thispagestyle{empty}

\subsection*{Abstract}
After a decade of research in low-power data collection, reaching arbitrary nodes has received comparatively little attention.
The leading protocol for low-power IPv6 routing, RPL, is no exception, as it is often studied in multipoint-to-point scenarios.
As a result, downward routing (from root to node) is still notoriously difficult, holding back the emergence of Internet of Things applications that involve actuation.
In this paper, we focus on achieving industrial-grade reliability levels (1e-5 failure rate) in downward routing with RPL.
We make every packet count, and classify the different causes of packet loss.
We show how to mitigate each source of packet loss, by (1) introducing a gradient metric that favors reliable links, (2) increasing neighborhood awareness for accurate link selection, and (3) ensuring a robust routing state maintenance and packet forwarding.
We demonstrate RPL downward routing with loss rates in the order of 1e-5 in four different testbeds up to 352~nodes, in both sparse and dense settings.
We also validate our solution on top of a low-power TSCH scheduler, and achieve sub-percent low duty cycles and a channel utilization of 0.07\% at every node, spread over 16~channels.

\section{Introduction}
\lsec{introduction}

As the field of low-power wireless communication matures, the focus started shifting from best-effort to reliable communication.
Industrial monitoring applications now routinely achieve at least {\em five nines} of reliability (99.999\%, or 1e-5 failure rate)~\cite{doherty07channelspecific,dombrowski15echoring}.
Such a level of reliability, if brought to the broader Internet of Things realm, would enable a whole new class of applications, where communication success is the norm.
Interactive systems, control loops and reliable file transfer would all benefit from a more robust infrastructure.

We focus on IETF's low-power IPv6 stack, in particular 6LoWPAN, RPL, and 6TiSCH, which is a leading set of standards for the interoperable Internet of Things (IoT).
Such low-power IPv6 networks are gaining momentum because of their openness, and application agnosticism.
They also bring addressable hosts, which caters for a broad range of scenarios, where nodes from different vendors can connect as IP hosts and interact via application-layer protocols on top.
Reliability is also critical for these applications, increasing user confidence in the technology.

After years of research and industrial advances, low-power data collection can now be done reliably, be it with RPL~\cite{duquennoy15orchestra} or other routing protocols~\cite{doherty07channelspecific,gnawali09ctp,duquennoy13orpl}.
Flooding-based solutions~\cite{ferrari12lwb,landsiedel2013chaos,istomin2016crystal} achieve a high reliability by eliminating the need for routing, but they are not ideally-suited for low-power IPv6, as discussed in~\rsec{discussion}.
Downward routing (from root to node), on the other hand, remains a challenge~\cite{bacceli11p2prpl,istomin15rplready}.
In RPL, downward routing with loss rates in the range of 1e-3 ({\em three nines}) has been reported~\cite{duquennoy15orchestra}.
While this may be sufficient for some applications, it is two orders of magnitude off the 1e-5 goal.

Reaching a failure rate of 1e-5 requires to make every packet loss count.
Through trace-driven simulations and large-scale experiments, we identify the different threats to packet delivery in RPL downward routing.
We confirm a number of known causes of loss such as link asymmetry and link probing issues, but also identify and tackle new ones, \eg the inherent reliability limitations of ETX or the inconsistent routing state problem.

We find that to enable robust routing, one needs a routing metric geared towards reliability and link symmetry rather than a best-effort metric such as ETX.
Second, we show that such a metric is of no use without an accurate view of the wireless environment at every node, and propose a link probing mechanism that allows RPL nodes to maintain fresh link estimates and keep the topology reliable at all times.
Third, once a reliable metric and probing are in place, routing state inconsistencies turn out to be the reliability bottleneck.
While this problem does not affect upward routing where nodes simply forward to their current parent, it is detrimental to downward routing.
We advocate source routing through RPL's non-storing mode as a mitigation strategy to this problem.
Finally, a factor that was negligible before becomes the main cause of packet loss: the IEEE~802.15.4 duplicate detection mechanism, based only on 8-bit sequence numbers, causes occasional spurious drops (at a rate of about 1e-4).
By eliminating this source of losses we achieve downward routing in several testbeds (sparse and dense, and up to 352~nodes) with a loss rate in the range of 1e-5 and below.

This paper makes the following contributions:
\begin{itemize}
\item We conduct an in-depth analysis of all causes of packet loss in RPL downward routing;
\item Via trace-driven simulations and testbed experiments, we characterize different relevant aspects, such as how link metrics affect reliability and asymmetry, or the degree of inconsistency in the routing state;
\item We present a set of reliability mechanisms that eliminate or mitigate all causes of packet losses by (1) gearing topology construction towards reliable downward routing and (2) increasing the consistency of both link estimates and the routing state;
\item We present a thorough experimental evaluation of RPL, where we demonstrate loss rates below 1e-5.
The evaluation involves four testbeds up to 352~nodes, both sparse and dense.
It includes results with a low-power TSCH scheduler.
We report on 850~one-hour experiments, one 2-week deployment, and in total over 12~million packets sent end-to-end.
\end{itemize}

\section{State of the Art}
\lsec{sota}

\paragraph{Low-power Wireless Networking}
Many protocols for low-power wireless networks build a gradient anchored at a root, typically used as the sink of a data collection network.
Examples of such include CTP~\cite{gnawali09ctp}, Dozer~\cite{burri07dozer}, ORW~\cite{landsiedel12low-power} or OppCast~\cite{mohammad16oppcast}.
Traditionally, the focus is energy-efficiency, latency, and reliability.
Reliability in the range of 90\% to 99\% is regarded as sufficient in a best-effort context.

The domain of industrial automation is more focused on guaranteed performance, with reliability as a major goal.
Solutions such as TSMP, WirelessHART or TSCH (part of IEEE802.15.4-2015~\cite{stdieee802154-2015}) enable solid low-power mesh networks with high reception rates~\cite{watteyne16industrialcps}.
For instance, Pottner~\etal achieved 99.95\% reliability in a 15-node testbed~\cite{pottner14constructing}, Pister~\etal reported 99.97\% to $>$99.99\% reliability~\cite{pister08tsmp}, while Doherty~\etal demonstrated 99.9995\% delivery over 49 nodes~\cite{doherty07channelspecific}.
The works above focus exclusively on data collection, while we tackle reliable routing to arbitrary nodes.

Data dissemination protocols such as Drip/CodeDrip~\cite{ribeiro14codedrip} or Pando~\cite{du15pando} were proposed, that disseminate a payload from the sink to all nodes.
These focus on efficient flooding, sometimes with network coding and other techniques, rather than on establishing one-to-one routes.
On the other hand, bulk transfer protocols such as Flush~\cite{kim07flush}, PIP~\cite{raman10pip} or BF~\cite{duquennoy11lossy} focus on reliable forwarding over an existing path; these are orthogonal to the routing protocol.

A radically different approach was introduced in 2011 with Glossy~\cite{ferrari11glossy}, a network flooding primitive that exploits constructive interference.
LWB~\cite{ferrari12lwb} was then designed to support multiple traffic patterns on top of Glossy.
The approach is appealing as it rules out the need for a routing protocol -- instead, all packets are sent repeatedly over the entire network.
LWB achieves delivery ratios in the range of~99.98\%.
Even higher delivery ratios were recently reported by Crystal~\cite{istomin2016crystal}.
These protocols, however, work best as integrated solutions, and are not geared towards IPv6 networks with generally larger payloads and interactive traffic patterns, unknown at deployment time.
They also result in an inherently higher channel utilization; a disadvantage when it comes to network co-existence.


\paragraph{Downward Routing in RPL}
RPL~\cite{winter12rpl}, the standard protocol for low-power IPv6, was built on ideas from CTP~\cite{gnawali09ctp}, extended with downward routing and other features.
The protocol builds a gradient for upward routing (node to root), and uses the reverse path to route downward (root to node) and reach any node.

RPL offers two distinct modes of operation: storing and non-storing mode.
In {\em storing mode}, nodes maintain routing tables locally, with enough information to reach their sub-tree.
Packets are routed upward until reaching an ancestor of the destination, where it engages downward routing.
In {\em non-storing mode}, the root maintains a complete view of the topology, but other nodes do not keep any routing state.
Packets are routed all the way up to the root, and then source-routed down to the destination.
This is done using an IPv6 source-routing header, which contains the path from root to destination (in a compressed form).

RPL was evaluated extensively~\cite{gaddour12rplnutshell}, with a focus on delay and throughput~\cite{accentura11rplperfanalysis} or the repair mechanism~\cite{tripathi10perfstudyrpl}.
Ko~\etal~\cite{ko11interop} proposed an interoperability analysis of TinyOS' and Contiki's RPL implementations.
The paper has key insights in achieving high performance in heterogeneous environments.
As far as reliability is concerned, in the aforementioned works, loss rates in the range of a few percent are generally reported.

The PRL link selection is done by children towards their parent.
When there is no upward traffic to update link estimates, or when links are asymmetric, the link from parent to child may be compromised.
This was observed experimentally on several occasions~\cite{ko2011rpltinyos,kim15marketnet}.
ORPL~\cite{duquennoy13orpl} proposed to use opportunistic routing in RPL, reaching downward reliability in the 99\% range.
Hydro~\cite{dawson10hydro} is conceptually close to RPL but with the ability to combine source routing with local routing tables.
Current extensions of RPL are also going in this direction, enabling the border router to remotely manage the nodes' routing table~\cite{thubert17daoprojection}.

A complementary approach in making RPL more reliable is to exploit more robust MAC layers.
MiCMAC~\cite{alnahas14micmac} applied channel hopping to low-power listening RPL networks, and reported delivery ratios of about~99\%.
Orchestra~\cite{duquennoy15orchestra} proposed a scheduling technique to exploit TSCH in RPL networks.
In a 98-node testbed, it achieved delivery ratios of 99.997\% in data collection, 99.98\% in downward routing.
As these are the highest delivery ratios reported on RPL so far, we select TSCH (with Orchestra) as the MAC layer for the experiments in this paper, both in traffic-intensive and power-saving scenarios.
Note that our focus is on the layers above MAC; TSCH is only used to enable testbed experimentation under realistic settings.

RPL has a number of limitations that were pointed out in the past, some of them tackled in this paper.
Iova~\etal{}~\cite{iova13stability} reported on the inherent trade-off between stability and link quality, which we further characterize and mitigate with our metric.
Dawans~\etal{}~\cite{dawans12link} discussed the problem of keeping link estimates accurate, which we address with probing.
An analysis of RPL by Clausen~\etal~\cite{clausen11critical} discusses, among other topics, link asymmetry and neighbor unreachability detection -- two problems we tackle in this paper.
Finally, in this paper, we evaluate a simple extension of ETX that prioritizes reliable links; other more advanced metrics~\cite{javaid10linkmetrics,karkazis12metrics} with various performance goals could be considered in future work.

To the best of our knowledge, this paper is the first to focus on reliability in downward routing, aiming for losses in the 1e-5 range (99.999\% delivery).
It characterizes in detail the different causes of loss, some of them already well-known, others not so much, and introduces new mitigation strategies.
We also carry out the largest public experimental validation of RPL to date, in four different testbeds, sparse and dense, and up to 352~nodes.

\section{Preliminary Study}
\lsec{preliminary}

\paragraph{Goal}
We start by characterizing the level of reliability achieved in a state-of-the-art implementation of RPL, in a large-scale testbed.
We instrument the communication stack to keep track of all packets with unique identifiers, and we log the journey of every single packet as it travels through layers and hops.
We then dissect the runs offline, count losses and assign them to different causes.

\paragraph{Setup}
For our study, we select Contiki~\cite{dunkels04contiki} since it features one of the most mature 6LoWPAN-RPL stacks~\cite{watteyne16industrialcps}.
We use the IoT-LAB testbed in Grenoble, with 352 M3 nodes equipped with a 2.4~GHz 802.15.4 radio~\cite{adjih15iotlab}.
We select node~240 as the root, in the middle of the deployment.
We run RPL in storing mode of operation, with Objective Function MRHOF and metric ETX (the default in Contiki).
The RPL topology built on top yields an average radius of 6.7~hops (distance to the node that is farthest from the root).

As the focus is on downward routing, we have the root send four packets per second to a randomly selected node.
We need such a high traffic load in order to collect enough data for statistically significant loss rates numbers (further discussed in \rsec{methodology}).
All experiments are run for one hour, and we always exclude the first 5~minutes to leave time for initial convergence.

We select TSCH as the MAC layer, as it enables high levels of reliability with RPL and is readily available in Contiki~\cite{duquennoy15orchestra,duquennoy17tsch}.
As the focus here is not on MAC scheduling or energy, we use the simplest possible TSCH schedule, which consists of a single shared slot in a slotframe of length~1.
Nodes basically wake up at every slot (10ms) to listen or contend for transmission.
We select such a schedule for its traffic capacity, as we need a high load in order to collect enough data points for statistical significance.
We evaluate low-power settings in \rsec{overhead}.

\tabpreliminary
\figsprobing

\paragraph{Results}
For the purpose of this preliminary study, we simply select the run that yields a median end-to-end Packet Delivery Ratio (PDR).
The losses in this particular run are summarized in~\rtab{preliminary}.
Overall, 99.3\% of the packets were successfully routed end-to-end, from root down to any node.
Out of 11,755 packets sent, 82~were lost, falling into three categories:
\begin{itemize}
\item \textbf{Cause~\#1 -- MAC-layer drop:} the main cause of loss is when the MAC layer discards a packet after 8~failed retries (default in Contki).
Note that the system is provisioned such as it does not saturate, that is, the MAC queues (of length 24) never get full.
\item \textbf{Cause~\#2 -- Route not found:} the next loss cause is routing inconsistency, \ie the packet reached a node that was not able to find a route to the destination.
\item \textbf{Cause~\#3 -- Spurious duplicate:} the last cause of loss is packets classified as duplicate while they are not, due the IEEE~802.15.4 1~byte sequence number.
\end{itemize}

In the remainder of the paper, we will show how to eliminate the second and third causes, and mitigate MAC-layer drops down to 0--1 loss per experiment.

\section{Design: Reliability Mechanisms}
\lsec{design}

We propose a number of mechanisms aimed at enhancing the reliability of the RPL mesh.
We address the causes of loss identified in \rsec{preliminary}, characterize the underlying problems through trace-driven simulation and testbed experiments, and introduce suitable mitigation mechanisms.

We first focus on link metrics, that is, what reliability can be achieved through optimal route selection, and what is the impact of asymmetric links.
This helps mitigate the first cause of loss, MAC-layer drops.
Second, we turn our attention to protocol aspects to make link estimation robust at runtime, and to ensure robust route dissemination and packet forwarding.
This eliminates the second and third causes of loss: inconsistent routing state and spurious duplicates.

\subsection{Link Metrics and Route Reliability}
\lsec{metrics}

\paragraph{Problem}
ETX~\cite{decouto05etx}, the metric traditionally used to build low-power mesh (\eg in RPL and CTP), is designed to optimize performance (throughput, latency, energy) rather than reliability.
For instance, a hop with 50\% PRR (ETX=2) will be deemed equivalent to two perfect hops (ETX=1+1=2).
When reliability is the first goal, the latter should be clearly preferred, as it yields an end-to-end delivery ratio of 100\% against 50\% for the former.

Formally, the end-to-end delivery ratio (PDR) for a given node $n$ is denoted:
$$
PDR(n) = \prod_{x \in path(n)}{{PRR_{x \rightarrow parent(x)}}^{1+R}}
$$
where $path(node)$ is the set of nodes in the path from $n$ to the root (including $n$ but excluding the root), $PRR_{x \rightarrow y}$ is the packet reception rate of the link $x \rightarrow y$, $parent(x)$ is the RPL parent of node $x$ and $R$ is the number of MAC retries.
The end-to-end loss rate is $LR(n) = 1-PDR(n)$.

We tackle the problem of selecting the best paths via parent selection in RPL, with the goal of minimizing the end-to-end loss rate (loss cause~\#1).

\figsasym

\paragraph{Approach}
We introduce a variant of ETX that balances path length and reliability.
The idea is to generalize ETX by raising the link cost (inverse of the PRR) at every hop to the power of a number N.
We denote the metric $ETX^N$.
The rank of node $n$ is computed as follows:
\begin{eqnarray*}
ETX^N(root) &=& 1\\
ETX^N(n) &=& ETX^N(parent(n)) + \\
 & & \left(\frac{1}{PRR_{n \rightarrow parent(n)}}\right)^N\\
\end{eqnarray*}

With $ETX^2$ for instance, the link cost at every hop is squared.
In the example above, the 50\% PRR link results in $ETX^2=2^2=4$, while the path with two perfect hops results in $ETX^2=1^2+1^2=2$.
Using $ETX^2$ as a metric, the path with two perfect hops will, therefore, be preferred.
Higher exponents (N) can be used to favor even more reliable links, at the cost of longer paths and more churn.
We also propose a metric that derives directly from our formulation of the end-to-end loss rate.
Nodes use their locally estimated $LR(n)$ as rank:
\begin{eqnarray*}
LR(root) &=& 0\\
LR(n) &=& 1-\left(1-LR(parent(n))\right)\\
      & & \times \left(1 - \left(1 - PRR_{n \rightarrow parent(n)}\right)^{1+R}\right)\\
\end{eqnarray*}
so as to select links that minimize their end-to-end loss rate to the root, regardless of the path length.
For instance, with two links at 50\% PRR, and assuming one retransmission, the nodes one/two hops away will respectively compute an LR of 75\%/56\%.

\paragraph{Characterization}
We use trace-driven simulations to characterize the reliability level theoretically achievable in a testbed for a given metric.
We first collect traces in the IoT-LAB testbed in Grenoble, over 352~m3 nodes.
We run the TSCH protocol at the MAC layer and assign a dedicated transmit slot to every single node in the network.
Nodes send a broadcast periodically (every minute) using their dedicated slot, and all packet receptions are logged.
We run the experiments for one hour.

We write an intentionally simple RPL simulator able to replay traces from the experiments.
At every transmission, the simulator updates every node's link estimate and rank and performs parent selection.
We compute the end-to-end loss rate of all upward paths (downward paths are dissected in \rsec{asymmetry}) assuming 8~link-layer retransmissions.
We compare the metrics ETX, ETX$^N$, and LR.

\rfig{probing} shows the results obtained with the different metrics (Tukey boxplot, per-node samples).
As expected, ETX$^N$ results in a higher end-to-end delivery ratio (\rfig{probing-pdr}), thanks to higher-quality links (\rfig{probing-prr}).
For instance, going from ETX to ETX$^2$ increases the worst link PRR from 51\% to 64\%.
The metric LR achieves the highest reliability by far, at the cost of an increased hop count (\rfig{probing-hops}) and churn (\rfig{probing-ps}).

Note that our attempts to use LR, ETX$^3$ or ETX$^4$ in a real system were unsuccessful due to the too long paths and high churn.
In particular, LR is very sensitive to slight changes in link quality partly due to its multiplicative nature.
This results in loops and prohibitively long paths.
Overall, we find the metric LR useful as an upper bound for reliability, while ETX$^2$ is more practical.
Also note that real values of N could also be considered, for even finer tuning.

\subsection{Link Asymmetry}
\lsec{asymmetry}

\paragraph{Problem}
Link asymmetry is a well-known problem in low-power wireless~\cite{chen09deal,sang07asymmetric}.
The success rates on the link $A \rightarrow B$ or $B \rightarrow A$ are often different, due to interference or fading.
In RPL, the problem is significant because the link between a child and its parent is selected by the child only, often based on upward link estimates.
The same link is, however, also used to relay downward traffic.
In the worst case of unidirectional links, a node might select a link that is not usable for downward routing at all, disconnecting the node and its subgraph.

There are many existing studies on the link asymmetry in real environments -- here, we investigate how different metrics affect asymmetry in RPL (asymmetry leads to packet drops, loss cause~\#1).

\paragraph{Approach}
We use the fact that, as demonstrated in previous work~\cite{sang07asymmetric}, good links are less prone to asymmetry than medium-quality or poor links.
We propose to utilize our metric ETX$^N$, which favors better links over the traditional ETX.
The idea is that the better links should also be more symmetric and hence more usable for RPL downward routing.

We formalize the downward route end-to-end PDR as:
$$
PDR_{down}(n) = \prod_{x \in path(n)}{{PRR_{parent(x) \rightarrow x}}^{1+R}}
$$
and the associated loss rate $LR_{down}(n) = 1 - PDR_{down}(n)$.

\paragraph{Characterization}
We use the same traces and simulator as in \rsec{metrics} but now extract the downward loss rate $LR_{down}$.
The routes are still built from child to parent, but we look at the resulting downward routes reliability.

\rfig{probing-downpdr} shows that the downward loss rates follow the same trend as for upward links: ETX$^N$ increases reliability and LR yields the best results.
Regardless of the metric, downward routes perform worse than upward routes, because routes are optimized upward and links are sometimes asymmetric.
\rfig{probing-downprr} shows the downward link PRR, and confirms our hypothesis: ETX$^N$, by preferring better upward links, also avoids asymmetric links.
The worst down link with ETX has a PRR of 28\%, while with ETX$^2$, all down links are above 50\% PRR.

\rfig{probing-parallel} offers a visualization of link asymmetry in the testbed.
Each sub-graph shows on the left the upward PRR and on the right the downward PRR.
The more horizontal the line, the more symmetric the link.
The first plot shows all links (in fact, only a representative 10\% subset), while the next ones show only the links selected by ETX and ETX$^2$.

First, notice how links can have any level of asymmetry, from perfectly symmetric to unidirectional (diagonal lines).
Fully unidirectional links are never selected by RPL because nodes need to hear their neighbors' beacon before using them as a parent.
ETX selects only a subset of links which are good upward, and usable downward albeit sometimes with a low PRR.
ETX$^2$ achieves two things: (1) it selects better links, shifting all lines up in the graph, and (2) it selects more symmetric links (as symmetry is correlated with link quality), leading to significantly improved down link quality.

\subsection{Link Probing}

\paragraph{Problem}
The discussions above on improving parent selection assume nodes have fresh link estimates for all their neighbors at all times.
By standard, RPL does not stipulate how to keep link estimates up to date -- it only mentions that some mechanism should take care of checking whether links are usable or not~\cite{winter12rpl}.
Without an accurate estimate of the link quality to its neighbor, a node is not able to get the best out of ETX or the other metrics discussed above, ETX$^N$ or LR.
Note that the link estimates are used for the upward link selection, but this directly affects downward routing, as discussed in \rsec{asymmetry}.
This problem contributes to loss cause~\#1.

\fignbrtable
\paragraph{Approach}
We propose to add a link probing mechanism to RPL in charge of keeping link estimates up-to-date.
The probing mechanism has three goals: (1) make sure our current parent is still reachable, (2) make sure we have a fresh link estimate for our backup parents so we can switch to one of them whenever the current parent fails, (3) investigate neighbors we once had a link to, so we can reuse the link when it is up again.

Our link estimator proceeds as follows:
\begin{itemize}
\item In addition to the link estimate, it keeps an exponentially decaying counter indicating metric freshness;
\item When first hearing from a neighbor, it exploits the incoming packet's RSSI to produce an initial guess PRR (inspired by the 4-bit link estimator~\cite{fonseca07fourble});
\item Periodically (we use a 1~min period), it selects a node for probing.
A RPL control message (unicast DIO) is sent and the link estimate will be updated from the resulting transmission count.
The neighbor selection function chooses the preferred parent if outdated, else it either (with 50\% likelihood) probes the best outdated potential parent or picks the least-recently updated neighbor.
This ensures all neighbors are eventually covered while keeping the best neighbors up-to-date at all times.
In addition, we schedule immediate probing whenever there is a need to switch to a parent with unfresh link estimate.
\end{itemize}

\paragraph{Characterization}
We run a 1h experiment with the same setup as our preliminary study (\rsec{preliminary}).
\rfig{nbrtable} shows the transmission count at node~\#23 towards its different neighbors.
Without probing, only a few neighbors are ever transmitted to, resulting in partial and often outdated link estimates.
With probing, on the other hand, more than half the nodes are covered.
This provides the node with more visibility on link estimates, resulting in a more educated guess for parent selection.

In sparser environments, probing is also helpful, as it increases the freshness of all link estimates.
For instance, in the 1h experiment, the top-4 nodes were all probed at least 6~times with probing, vs. only 1~time without.
Probing creates traffic overhead -- the cost of probing is evaluated in \rsec{mechanisms-contribution}.

\subsection{Routing State Consistency}
\lsec{routing-state-consistency}

\paragraph{Problem}
After selecting a preferred parent, nodes must spread this information up in the tree in order to become reachable.
In this process, the routing state may become outdated or worse, inconsistent (this is loss cause~\#2).

The RPL storing mode of operation, which is the most represented in the research literature, is prone to such state inconsistency.
In storing mode, each node maintains its own routing table -- but keeping consistent such a distributed routing state is hard.
For instance, assume a node N loses the link to its current parent A and switches to~B.
To update all routing tables, it has to de-register from A (and its ancestors) and register a new route to B (and its ancestors).
Should any of the updates be lost, the distributed state will become inconsistent.
For instance, in the absence of a link to A, the de-registration will fail.
As a result, all traffic to N traveling via A will be routed directly to N over the broken link, while it should now be routed upward to other nodes able to reach~N.

\paragraph{Approach}
We advocate source routing as a mitigation to routing state inconsistency.
In source routing, all state is centralized at the root.
When a node switches parent, all it needs to do is to keep the root informed about its new parent.
The loss of such information only results in an outdated state at the root -- but the root is guaranteed to have a consistent global view of the topology, where there exists a route to every node at any time (as it can reject or postpone any update that would introduce a transient loop).
RPL has a non-storing mode of operation which is based on source routing and is primarily designed for constrained nodes where storing routes is not an option.

We suggest selecting non-storing mode regardless of the nodes' memory constraints, whenever the reliability of downward routes in crucial.
For this paper, we contribute an implementation of non-storing mode and evaluate it systematically against storing mode.

\figstopo
\paragraph{Characterization}
We run an experiment in the testbed, letting RPL construct a 352-node topology over the course of 1h.
We repeat this for both storing and non-storing modes of operation.
We use the globally synchronized TSCH clock to trigger logging of the routing state (parent and routes) at every node, and we process the logs offline to analyze the routing state consistency.

\rfig{topology-st} shows the evolution of the routing state in storing mode.
After initial convergence, all nodes join the topology and become reachable.
However, throughout the experiment, a portion of the nodes (about 50) suffer from an outdated state.
This means that at least one node in the network has a missing or stale route to that destination.
Occasionally, a few (one or two in this run) nodes become transiently unreachable.
This means that a packet starting at the root would not find a route to the node -- it would be discarded somewhere along the path.

In contrast, \rfig{topology-ns}, shows the routing state consistency in non-storing mode.
After the initial convergence, no node ever becomes unreachable again.
The occurrences of an outdated state are also much rarer than in storing mode: whenever an update is lost, the next update is sufficient to restore a correct state at the root.

\tabsummary
\subsection{Duplicate Detection}

\paragraph{Problem}
Packets in IEEE~802.15.4 use a 1-byte sequence number for MAC-layer duplicate detection.
With only a single byte to discriminate packets, however, false positives may arise (loss cause~\#3).
For instance, assume node~A transmits a packet to B with sequence~0, then it transmits 255~more packets to other nodes, and then another packet to B after the sequence has wrapped to~0.
Node~B will assume the second reception is a duplicate and drop the packet.
We observed this very scenario in our traces, as well as more convoluted ones also leading to spurious duplicate detection (see~\rtab{preliminary}).

\paragraph{Approach}
We take a number of steps to eliminate spurious duplicate detection.
First, we propose to use the sequence number suppression feature of IEEE~802.15.4 frames v2 for all broadcast frames.
This essentially reduces the pace at which sequence numbers wrap.
This practice has no downside as far as we know, as broadcasts are not re-sent and do not suffer from duplicates.

Second, we store only the last sequence number for every neighbor, instead of a list of all recent packets seen.
This is unlike what mainline Contiki does, but still standard-compliant and comes with no drawback as long as sequence numbers are suppressed from all broadcast.

Third, we add a lifetime to sequence numbers.
In the duplicate detection table at each node, we add information about the reception time.
After a fixed lifetime (we use 30~seconds), we simply expire the sequence number.

Lastly, we extend link-layer acknowledgments to include the MAC address of the sender for the packet being ACKed (standard option of frames v2).
This solves a related problem we encountered occasionally, where two nodes send a packet with an identical sequence number in the same slot.
In this case, ACKs without a MAC address are ambiguous, \ie they are not enough for the sender to know which of the two packets it being ACKed.

\paragraph{Characterization}
This solution results in an 8-byte overhead on ACK frames (sender MAC address), saves 1~byte in every broadcast frame (sequence number suppression), and results in lower RAM footprint (single sequence number stored instead of a list).

\section{Evaluation}
\lsec{eval}

This section assesses the mechanisms proposed in this paper, with a focus on reliability and overhead.

\figscomparison
\subsection{Methodology}
\lsec{methodology}

We implement non-storing mode and our reliability mechanisms in RPL for Contiki.
We make our source code available at \url{http://double-blind.review}.

\paragraph{Scenario}
We use a setup similar to that of our preliminary study in ~\rsec{preliminary}.
The root sends a packet to any node at 4~Hz.
The UDP payload is 16~bytes, and all datagrams could fit single packets (no fragmentation).
At the MAC layer, we use TSCH (hopping over all 16~channels) with a simple schedule where all nodes wake up at all slots to listen or transmit.
Unless otherwise mentioned, all experiments are one hour long and we start recording metrics first after five minutes to account for initial convergence.

\paragraph{Statistical Significance}
As we intend to measure loss rates in the order of 1e-5 and below, we need to gather a large number of data points for statistical significance.
Our settings (4~Hz packet interval and traffic-intensive TSCH schedule) are designed precisely to maximize the number of packets in each experiment.
More realistic scenarios, with a low power scheduler and a lower traffic load, are evaluated in~\rsec{overhead}.

With a packet interval of 4~Hz, each run involves about 11,700 application packets sent, end-to-end.
Each experiment is run at least 10~times.
For statistical significance, more runs are added until we reach at least 10~losses or totaling over half a million end-to-end transmissions (takes about 50~iterations).
When reaching half a million points with no single loss, \emph{the rule of three}~\cite{eypasch95ruleofthree} (for statistical significance of zero-event samples) states that the loss rate is at most 6e-6, with 95\% confidence.
A summary of our experiments, with loss rates and packet count in each setup, is shown in \rtab{summary}.

\paragraph{Metrics}
We focus on the following metrics:
\begin{itemize}
\item \emph{End-to-end loss rate:} application packets loss rate, with link-layer retransmissions and over multi-hop;
\item \emph{Link PRR:} link packet reception rate, before retries;
\item \emph{Latency:} the end-to-end delay for application packets;
\item \emph{Duty cycle:} the portion of time spent with radio on;
\item \emph{Channel utilization:} the portion of time spent with the radio transmitting.
\end{itemize}

We use four different testbeds (in a total of five settings), presented next and with detailed properties summarized in \rtab{summary}.
We use the IoT-LAB testbeds~\cite{adjih15iotlab} in Grenoble (largest testbed) and Lille (densest testbed).
Experiments run on a selected subset of 52~nodes in Grenoble, ``Grenoble-52'' allow to evaluate sparser scenarios.
We use Flocklab with OpenMotes for smaller-scale settings, and JN-IoT, a private testbed that spans half of a floor in a research building.


\subsection{Contribution of Each Mechanism}
\lsec{mechanisms-contribution}

We first look at the individual contribution of the mechanisms introduced in \rsec{design}.

\paragraph{Setup}
We select the IoT-LAB Grenoble site for its challenging scale (352~nodes), and use our 4~Hz downward routing scenario.
We run RPL in both storing and non-storing modes, with none of our mechanisms (baseline), and then incorporate our reliability mechanisms.

\paragraph{Results: Reliability}
\rfig{comparison-pdr} shows the end-to-end loss rate obtained in each scenario.
The baseline, without our mechanisms, yields a loss rate in the \num{10e-2} range, that is, two nines of reliability (PDR of 99\%).
Probing gives us one order of magnitude improvement, reaching below \num{10e-3}.
This is the result of a better link estimation and selection, as confirmed by the link PRR shown in \rfig{comparison-prr}.

Our enhanced duplicate detection eliminates all spurious duplicates, which were responsible for \num{10e-4} to \num{10e-3} of losses (as can be seen in \rtab{summary}).

Our metric $ETX^2$ enables non-storing mode to lower its loss rate by another order of magnitude, reaching \num{10e-4}.
This is the result of prioritizing high-quality links (median above 90\%, see \rfig{comparison-prr}).
The metric, however, has a negative impact on RPL in storing mode.
This is because of the increased hop count and churn, which yields more routing inconsistencies (\cf \rtab{summary}).

We explore additional configuration options in order to push the reliability envelope (we use the most reliable metric, \ie $ETX$ for storing and $ETX^2$ for non-storing).
First, we switch from using all 16~channels to only the 4~channels that are least affected by Wifi (channels 15, 20, 25 and 26).
This leads to a direct increase of the link PRRs (\rfig{comparison-prr}).
Second, we increase the number of MAC retries, from 8 to 16 and then 32 (still on Wifi-free channels).
The latter setting allows us to gain one more order of magnitude and reach, in non-storing mode, a loss rate of \num{8e-6}, that is, a PDR over 99.999\%.

\paragraph{Results: Latency}
We now look in \rfig{comparison-latency} at the impact of our mechanisms on the end-to-end latency.
There is one data point per node and per run.
In all experiments, the mean latency is below 0.11~s.
With our mechanisms enabled, it ranges between 0.05 and 0.07~s.
As we incorporate our reliability mechanisms, we eliminate more and more outliers.
Notice how increased MAC retries, in contrast, re-introduces outliers, due to some packets delivered after a large number of attempts.
In our experiments with 32~retries (over 1.3M data points in total), the 99\textsuperscript{th} percentile packet latency is 0.27~s.

\tabatoa

\subsection{Performance in Different Testbeds}

We now assess our reliability mechanisms in testbeds with different physical topologies, resulting in various logical depth and density.

\paragraph{Setup}
We now run RPL in non-storing mode with and without our reliability mechanisms, in the IoT-LAB deployments in Grenoble-52, Lille, Flocklab, as well as in JN-IoT.
We run the same application as above, that is, 4~Hz downward traffic.
We use the best-suited metric for each mode-of-operation, that is, ETX in storing mode and $ETX^2$ in non-storing.

In the Lille testbed, where all the nodes are located in the same room, we use the minimum transmission power: -17dBm.
At -17dBm, nodes can still hear each other, but the decreased link quality forces RPL into selecting multi-hop paths.
To reduce the traffic load, we set two third ($2/3$) of the nodes to run as leaf nodes, and change the application traffic rate from 4~Hz to 1~Hz.

\paragraph{Results}
\rtab{summary} summarizes the results.
The results tagged \emph{Orchestra} are with a low-power scheduler and are discussed in \rsec{overhead}.
In all testbeds, our mechanisms improve the loss rate by one or two orders of magnitude, reaching 2e-5 and below.
In Grenoble-52 and Lille with the most reliable settings, we did not observe any loss out of over 500K end-to-end transmissions.
Interestingly, we did not notice any spurious duplicates in other testbeds than Grenoble -- a fact that we attribute to the less challenging topologies, with fewer nodes and hops.

This series of experiments demonstrates the usefulness of our various reliability mechanisms in both sparse and dense testbeds, at both large and small scale.
The measured radius (mean hop count of the farthest away node) ranges between 2.4 and 6.7, while the density (mean number of neighbors) is between 5.4 and 237.

\figsorchestra
\figtraffic

\subsection{Point-to-point Routing}
\lsec{p2prouting}

After our initial focus on downward routing, we investigate how our mechanisms help in a more general traffic pattern, where any node sends to any node.
This involves routing upward first (to the root in non-storing mode, or to any common ancestor in storing mode), and then downward to the destination.

\paragraph{Setup}
We focus again on the largest-scale testbed at our disposal: the IoT-LAB Grenoble site.
We select 10\% of the 352~nodes (uniformly distributed) as data sources.
Each source sends data at a 20-second interval, with jitter.
The destination is selected at random, but nodes use the same random seed across experiments, ensuring a deterministic selection of destinations.
We only run RPL with our reliability mechanisms enabled, in either storing or non-storing mode, with either ETX or ETX$^2$.

\paragraph{Results}
\rtab{atoa} summarizes the point-to-point routing experiments.
In storing mode, the first cause of packet loss is routing inconsistencies.
In these experiments, non-storing mode improves reliability by two orders of magnitude.
It reaches a loss rate of 1e-5 (PDR: 99.999\%), vs. 5e-3 for storing mode (PDR: 99.5\%).

On the other hand, storing mode is superior when it comes to path length and latency.
Depending on the settings, 13--23\% of the traffic was routed point-to-point without involving the root, resulting in on average 12--26\% shorter latency.
Nodes benefit from such shortcuts whenever they belong to the same logical sub-tree.
Note that non-storing also has a fraction of its traffic routed without the root; this happens whenever the destination is on the path up from source to root.
\rsec{discussion} discusses further the strengths and weaknesses of both modes of operation.


\figlongtimeline

\subsection{Energy-efficiency and Overhead}
\lsec{overhead}

This subsection quantifies the energy-efficiency and traffic overhead of our mechanisms.

\paragraph{Low-power Operation}
We first characterize our solution with all mechanisms enabled.
We use the Orchestra scheduler~\cite{duquennoy15orchestra} for TSCH, which lets nodes maintain their slots autonomously to match the current RPL topology.
Orchestra provides different slotframes to support synchronization traffic, rendezvous, and unicast communication.
The slotframes must have lengths that are co-prime, such as they overlap in a uniformly distributed way.
We dimension Orchestra as follows: the synchronization slotframe has a length of 383 (one transmit slot and one receive slot that repeat every 3.83~s), and the rendezvous slotframe a length of 101 (single slot that repeats every 1.01~s).
We vary the unicast slotframe length between~3 and~59 (one transmit slot and one receive slot that repeat every 30 to 590~ms).

\rtab{summary} reports on the reliability obtained using Orchestra with a unicast slotframe of length~7 and a packet interval of 4~Hz.
Out of over 606K packets, we measured a loss rate of 3e-5.
Collecting statistically significant numbers for longer slotframe lengths turned out challenging, due to the reduced network capacity.
The slotframe length has in theory no reason to affect reliability, as long as the traffic load is below the capacity of the schedule.

\rfig{orchestra-latency} shows the latency as a function of the unicast slotframe length.
We decrease the packet rate to 1~Hz so as not to saturate in the case of long slotframes.
We measure a latency proportional to the slotframe length, ranging from 0.1~s to 1.5~s.
The radio duty cycle (\rfig{orchestra-dc}), on the other hand, is inversely proportional to the slotframe length.
At a slotframe length of 29, the latency is 0.61~s with a duty cycle of 0.44\%.

\rfig{orchestra-dctx} shows the channel utilization at every node (\ie the duty cycle of transmit mode).
The full systems has a noticeably low channel utilization, below 0.07\% in all our runs.
Further, note that the utilization is spread over all 16~channels, \ie, the average per channel is below 0.005\%.
This is enabled by TSCH, where nodes wake-up synchronously to transmit/receive single packets before going back to sleep.




\paragraph{Overhead of the Reliability Mechanisms}
Our mechanisms showed no noticeable effect on the system's duty cycle, due to the Orchestra schedule being dominated by idle listening, with comparatively negligible channel utilization (\rfig{orchestra-dc} and \rfig{orchestra-dctx}).
To characterize the cost of the reliability mechanisms fairly, and independently of a given scheduler, we instead look at the traffic load.

\rfig{traffic} shows the amount of MAC-layer traffic, that is, all transmissions and retries, that take place at every node in Grenoble (4~Hz packets).
In total, the more reliable configurations cost 27\% extra traffic in storing, and 15\% in non-storing mode.
The main source of overhead is unicast DIOs, used for our probing mechanism, adding about 1.5~packet per minute.
Our reliability mechanisms, on the other hand, result in fewer retransmissions (see "MAC retries") through the selection of stronger links.

Further, note that non-storing mode inserts a source routing header in every packet routed downward.
Specifically, the RPL/IPv6 source routing header takes 6~bytes, followed by compressed addresses of all nodes in the path.
The network prefix (8~bytes) is common to all nodes and hence not included.
The device IDs (next 8~bytes) are compressed dynamically, such as bytes shared by all devices in the path are not repeated.
In all testbeds with experimented with, addresses could be compressed down to 2--4~bytes each.
In deployments with heterogeneous nodes, MAC addresses may differ widely, resulting in a compressed form up to 8~bytes.

\subsection{Long Experiment}

We finally assess our reliability mechanisms in the long run, over the course of a two-week experiment.

\paragraph{Setup}
We use the downward-traffic scenario with a traffic rate at 4~Hz, in the JN-IoT testbed.
We let the application run for two weeks, for a total of 4.8~million packets.
We use the WiFi-free channels and 32~MAC retries.

\paragraph{Results}
\rfig{longtimeline} shows a timeline covering the whole duration of the experiment, with time split into 1-hour chunks.
The top two graphs show end-to-end losses, split into MAC drops and queue overflows; the middle graph shows link PRRs; and the bottom graph parent switches.

In total, 444~losses occurred out of 5.1~million packets, \ie an average loss rate of 9e-5.
427~losses are due to a queue overflow (8e-5) and 16~to a MAC drop (3e-6).
There are two peaks of losses, dissected next.

The first peak (shortly before day~8) is correlated with an overall link degradation in the testbed (noticeable from the lower link PRR and resulting parent switches).
Over the course of 140~minutes, packets accumulate in the queue (of length~24) as more retransmissions are needed, leading to 77~queue drops.

The second peak (shortly before day~10) is triggered by the severe degradation of the link between the root and node \#10, which was on the path to several other nodes in the network.
The link is degraded for about 8~minutes, and some packets are dropped after 32~retries.
In our settings (TSCH min and max backoff exponent resp. 1 and 7), a packet can spend up to 35~s in a queue; enough time for the root to enqueue 140 additional packets (4~Hz send period).
By the time node \#10 notices the link conditions and decides to switch parent, many more packets are added to the queue.
With such a high pressure on the queue, 328~overflows occur.

Note that on a less traffic-intensive scenario or with a larger packet queue (external storage could be used as a backup), the queue overflows could be avoided -- at least for transient link degradations such as witnessed in this experiment.
In all cases, the network was back to normal performance once the link qualities increased again.

Finally, note the high parent switch count overall, with about 4~switches per hour for every node.
This is something we noticed only after the fact, and we can attribute to a too low rank-hysteresis threshold for this testbed.
It is worth noting that although this parent switching rate seems far from optimal, it was not detrimental to reliable operation, in part because non-storing mode always keeps a consistent state at the root even in the occurrence of high churn, as seen in \rsec{routing-state-consistency}.

\section{Discussion}
\lsec{discussion}

We discuss here aspects such the generality and limitations of our work.

\paragraph{Is an extra 0.1\% delivery worth the overhead?}
Going from 99.9\% to 99.999\% is a $100\times$ loss reduction -- which we achieve at a cost up to 27\% extra traffic (\rsec{overhead}).
This enables applications with five-nines reliability, even without end-to-end retransmissions.
Applications that can not tolerate any loss will see the end-to-end retransmission load decrease.
Best-effort traffic (\eg network management with ICMP) will also benefit from higher delivery ratios, improving quality of experience.

\paragraph{Why 4~Hz downward traffic?}
Our traffic pattern is tailored for statistical significance, so as to generate millions of data points (see \rsec{methodology}) .
As the mechanisms are independent of the traffic load, they are equality beneficial in lower traffic scenarios (higher traffic loads, on the other hand, could saturate the network).
Finally, upwards traffic (partly evaluated in \rsec{p2prouting}) should also benefit from $ETX^N$ and link probing.

\paragraph{Is storing mode still useful at all?}
Although we find storing mode be less reliable, non-storing mode is no silver bullet.
The source routing headers can trigger fragmentation of larger datagrams, especially in networks with many hops.
Further, for nodes belonging to the same sub-graph (13--23\% of all traffic in our \rsec{p2prouting} experiments), storing mode can find root-free routes and achieve lower latency.

\section{Conclusion}
\lsec{conclusion}

This paper addresses the challenge of reaching arbitrary nodes reliably in RPL.
We systemically identify the causes of packet loss and mitigate them with new reliability mechanisms.
Through an extensive testbed evaluation campaign, we demonstrate loss rates below 1e-5 at a reasonable overhead (15--27\% extra traffic).
We believe that our approach and findings can benefit not only RPL but other low-power routing protocols such as P2P-RPL, LoadNG or RIPng, and enable new applications that build on reliable low-power IPv6 routing.

{\footnotesize \bibliographystyle{acm}
\bibliography{compact}}

\end{document}